\documentclass[manuscript,screen]{acmart}

\AtBeginDocument{%
  \providecommand\BibTeX{{%
    \normalfont B\kern-0.5em{\scshape i\kern-0.25em b}\kern-0.8em\TeX}}}

\setcopyright{cc}
\setcctype{by}
\acmJournal{CSUR}
\acmYear{2025} \acmVolume{1} \acmNumber{} \acmArticle{1} \acmMonth{8} \acmPrice{}\acmDOI{10.1145/3759914}

\begin{document}

\title{Boosting Mixed-Initiative Co-Creativity in Game Design: A Tutorial}

\author{Solange Margarido}
\email{solange@dei.uc.pt}
\orcid{0000-0002-8039-7188}
\author{Penousal Machado}
\email{machado@dei.uc.pt}
\orcid{0000-0002-6308-6484}
\author{Licínio Roque}
\email{lir@dei.uc.pt}
\orcid{0000-0002-1911-2788}
\author{Pedro Martins}
\email{pjmm@dei.uc.pt}
\orcid{0000-0002-3630-7034}
\affiliation{%
  \institution{University of Coimbra, CISUC/LASI, Department of Informatics Engineering}
  \city{Coimbra}
  \country{Portugal}
}

\renewcommand{\shortauthors}{S. Margarido et al.}

\begin{abstract}
  In recent years, there has been a growing application of mixed-initiative co-creative approaches in the creation of video games. The rapid advances in the capabilities of artificial intelligence (AI) systems further propel creative collaboration between humans and computational agents. In this tutorial, we present guidelines for researchers and practitioners to develop game design tools with a high degree of mixed-initiative co-creativity (MI-CCy). We begin by reviewing a selection of current works that will serve as case studies and categorize them by the type of game content they address. We introduce the MI-CCy Quantifier, a framework that can be used by researchers and developers to assess co-creative tools on their level of MI-CCy through a visual scheme of quantifiable criteria scales. We demonstrate the usage of the MI-CCy Quantifier by applying it to the selected works. This analysis enabled us to discern prevalent patterns within these tools, as well as features that contribute to a higher level of MI-CCy. We highlight current gaps in MI-CCy approaches within game design, which we propose as pivotal aspects to tackle in the development of forthcoming approaches.
\end{abstract}




\maketitle

\section{Introduction}

Mixed-initiative and co-creativity are two increasingly prominent subjects within the research areas of human-computer interaction and computational creativity. As AI systems are becoming increasingly powerful and sophisticated, we are shifting from the view of AI as a tool towards AI as an active participant. Thus, there is a growing concern among researchers in exploring how we can benefit from human-AI partnerships. Particularly, there has been extensive research into the interaction between humans and computational systems to foster a collaborative environment where both agents exhibit initiative and proactivity, hopefully enhancing the creative process \cite{Yannakakis2014}. As stated by Davis\cite{Davis2013}, ``human-computer co-creativity describes a situation in which the human and the computer improvise in real-time to generate a creative product''. 
Mixed-initiative systems were defined by Carbonell \cite{Carbornell1970} as those ``in which both humans and machines can make contributions to a problem solution, often without being asked explicitly''. More recently, Yannakakis et al. \cite{Yannakakis2014} introduced the term \textit{mixed-initiative co-creation}, describing it as ``the task of creating artifacts via the interaction of a human initiative and a computational initiative'', emphasizing the proactivity of both parties. Although the use of different terms to describe mixed-initiative may cause some ambiguity, the addition of \textit{co-creativity}, or an alternative term such as \textit{human-computer co-creativity}, can serve as a highlight of the close collaboration among the agents and the creative nature of the process and its resulting output. Throughout this paper, we refer to \emph{mixed-initiative co-creativity} (MI-CCy) as we are focusing on a strong creative partnership between humans and computers.

Video games are a domain of particular interest for the application of mixed-initiative co-creative approaches. As the market grows and the players' expectations increase, so does the time, money, and effort needed to create games with large amounts of unique content that meet the expected quality \cite{Shaker2016,Baldwin2017, Walton2020}. For this reason, several support tools (e.g., level editors or game engines) have been developed to streamline game development processes \cite{Yannakakis2014}. Along with these, there has been major research and development of tools to automate content generation, known as procedural content generation (PCG) \cite{Shaker2016,Baldwin2017}. While very beneficial, there is no human-computer co-creation in the use of these tools. When using support frameworks, the creative process relies merely on human initiative. As for PCG, the human creative input is very limited, mainly consisting of the choice of parameters before and/or after the generative process \cite{Yannakakis2014}. The designer's lack of control often results in a lower quality of user experience, with generated content coming off as uninteresting and predictable \cite{Karavolos2015, Shaker2016}. Mixed-initiative approaches applied to game design have the potential to address these issues, as humans and machines work together, both proactively contributing to the creative process.

In this tutorial, we provide directions for researchers, developers, and designers to effectively incorporate a high level of mixed-initiative co-creativity into their work within the game design domain. Readers will gain an understanding of the characteristics and shortcomings of existing mixed-initiative co-creative approaches for various types of game content. They will also learn about the key elements that foster a balanced partnership and seamless collaboration between agents and how to identify and analyze these elements within a mixed-initiative co-creative tool. In Section~\ref{classes}, we establish a categorization for game content. We follow in Section~\ref{MIGameDesign} with an overview of selected works for each class of game content, summarizing their features and how they foster co-creativity between human and computational designers. In 
Section~\ref{MIclassification}, we introduce a framework for assessing the level of MI-CCy using rating scales for each criterion, and we analyze the selected works through the use of this framework. In Section ~\ref{Discussion}, we discuss the effectiveness and impact of these solutions while identifying some open issues. Finally, we conclude the survey in the last section.

\section{Classes of Game Content}\label{classes}
Game design is about deciding everything that a game should be, that is, everything that the player experiences \cite{Schell2008}. As this concept is very broad, we must break it down. Therefore, to allow for categorization of the works presented in Section~\ref{MIGameDesign} by the tackled game design domains, we present six classes of game content, starting from simple and concrete units to more complex and conceptual aspects. These classes were defined based on the classes of procedurally generated game content presented by Hendrikx et al. \cite{Hendrikx2013}, as it can closely apply to game content in general. Some changes were made to include a greater amount of content that was not mentioned by Hendrikx et al., such as objects, characters, physics, mechanics, and user interface. Furthermore, we strictly focus on in-game content, excluding derived content such as news and broadcasts of games.

\paragraph{Game Bits}
Game bits are ``elementary units of game content'' \cite{Hendrikx2013}, in which, like Hendrikx et al. \cite{Hendrikx2013}, we include textures, sound, buildings, and elements of nature such as vegetation, fire, water, stone, and clouds. We also include 2D image backgrounds, objects of any kind (2D or 3D representations), characters (only their visual representation, not their personality and behavior), and user interface elements such as buttons. These elements can be in their look and shape, or they may still be in a sketch state.

\paragraph{Game Space}
The game space represents the environment in which the game unfolds. It can range from a preliminary outline of the spatial structure, the components it may contain, and their respective placements, to a high-level representation, complete with the final look and all the associated game bits in place. When talking about game space, we mainly refer to maps, which can be indoor or outdoor and are composed of navigable space, obstacles (e.g., walls, rocks, and rivers), and interactive elements, such as collectibles. 
Some common map types include dungeons, consisting of a group of rooms interconnected by corridors containing enemies and treasures \cite{Dahlskog2015}, and platform games, which usually feature suspended platforms with varying shapes and sizes placed at different heights.

\paragraph{Game Behaviors}
In this tutorial, we define the game behaviors class as the predefined way game elements, such as objects and characters, act and interact with their surroundings. This includes their relationship with the physics of the game (e.g., whether they float or react to gravity) and their behavior in different kinds of interactions such as moving, expanding, exploding, and disintegrating. It also includes the various mechanics of each game element and their specific characteristics (e.g., jumping, running, throwing, and sliding). The characters' own way of speaking and acting and their personality traits are also associated with game behaviors. 

\paragraph{Game Systems}
Game systems combine sets of rules and algorithms to provide and manage complex (and often sustainable) interactions between game elements at a large scale. Hendrikx et al. \cite{Hendrikx2013} identify four main types of game systems: ecosystems, which dictate the evolution and interplay of plant and animal life; road networks, which affect traffic, resource exchanges, and access to services in outdoor locations; urban environments, concerning the efficient organization of clusters of buildings for the resident population; and entity behavior, such as complex interactions of non-playable characters with the player and as a group.

\paragraph{Game Scenarios}
Game scenarios outline the sequence of game events and how they unfold \cite{Hendrikx2013}. In earlier stages of game design, game scenarios can be portrayed through media such as storyboards or diagrams. The story of a game is a prime example of a game scenario, primarily prominent in narrative-driven games. Puzzles provide a space of possible solutions that may require the occurrence of very specific orders of events. Moreover, this class encompasses the strategic organization of game levels and the careful planning of level progression in accordance with an underlying mission or set of objectives.

\paragraph{Game Design}
Finally, we must refer to the most general and conceptual aspects of a game, such as rules, goals, genre, aesthetic theme, historical context, depicted universe, and narrative tone. Converting these concepts and ideas into a game involves exploring the various classes of content mentioned above. In fact, a game can be designed in our head \cite{Schell2008}, but by going through the myriad steps of creating game content, we can turn it into a physically playable artifact.

\section{Mixed-Initiative in Game Design: Literature Review}\label{MIGameDesign}

This section gathers a selection of current works involving the development and exploration of mixed-initiative co-creative approaches for game design processes. The selection made was based on the search for the keywords ``mixed-initiative'' along with ``game design'', ``game'' or ``games'' in Google Scholar, focusing on relevance and topicality. The same search combinations were made by replacing ``mixed-initiative'' with ``human-computer co-creativity'', ``human-computer co-creation'', ``human-AI co-creativity'', ``human-AI co-creation'', and ``co-creative AI'' so that no work would be missed due to the use of different terms to express the same idea. The focus on these terminologies ensures that the selected tools not only involve the use of AI for creative tasks in game design but also provide a clear description of how human and computational agents interact and collaborate creatively. 

To keep the focus on games, the selection of works we conducted prioritized mixed-initiative approaches exclusively applied to game design contexts. Although we did find some studies that could potentially have application in game design but did not primarily target it, such as mixed-initiative approaches to writing poetry \cite{Oliveira2017, Chakrabarty2022} or creating narratives \cite{Samuel2016, Stefnisson2018, Kreminski2022, Yuan2022, Lee2022}, we chose not to include them in our case studies, as they are not tailored to the specific needs and structures of game development. Additionally, our selection focuses exclusively on video games and does not include related media such as board games or tabletop role-playing games. While AI applications in these domains are emerging \cite{Triyason2023}, especially through general-purpose technologies like ChatGPT, systems specifically designed to support mixed-initiative co-creativity in these contexts are still scarce. Our focus on video games reflects both the current state of the art and our interest in purpose-built game design tools.

We decided to narrow down our selection of works to ten candidates in order to have a sample size that is sufficiently small to allow for a detailed analysis yet large enough to be representative of the state of the art. We aim to include at least one example for each class of game content. For those classes that are more prevalent in MI-CCy systems, we included more than one exemplary work, reflecting the prevailing trends in the field. These classes are game space, with a particular focus on map design, followed by game scenarios, which places greater emphasis on narrative design, and then game bits. The game systems class does not have any work selected to be analyzed with our framework due to the lack of literature that sufficiently meets our MI-CCy criteria.
In determining which works to include for each game content class, our primary criterion was to feature recent works, as well as works that held significant relevance in this research area. We sought to ensure that there were substantial differences among the various candidates to guarantee diversity in our study. Another crucial criterion in the selection process was the availability of the system online, through a web application or repository, for testing purposes. The direct interaction with the systems allowed us to conduct a more in-depth analysis.

It is worth noting that our selection does not include systems from the commercial sector. This gap is due to two key factors: first, the lack of public and free access to most commercial tools, which prevents direct evaluation; and second, our aim to maintain a focused scope on tools that explicitly support mixed-initiative co-creativity in the context of game design. While there are notable commercial tools that apply AI to creative processes, to the best of our knowledge, most are either not specifically tailored for game design (e.g., KREA \cite{RodriguezPrado2024}) or do not meet the criteria for MI-CCy or provide sufficient detail about the human–AI collaboration to be meaningfully analyzed within our framework (e.g., Popul8 \cite{Dias2024}). That said, we recognize that commercial developments in this area are highly relevant and contribute significantly to advancing AI-supported creativity in practice.
The final candidates are presented below, grouped according to the class of game content most prevalently addressed by each approach.

\subsection{Game Bits}
Mixed-initiative tools specifically aimed at creating game bits are scarce in the literature. There are several co-creative systems for art creation; however, they can easily be used to produce game content. This is the case of systems for co-creative sketching, such as Creative Sketching Partner \cite{Karimi2020}, where a user starts drawing and can ask the artificial agent for inspiration at any time. The system responds with a sketch with a desired amount of visual and conceptual similarity to the user's sketch. Then, the user may be inspired to make changes to their sketch and repeat the process if they wish to. Despite the diversity of mixed-initiative tools potentially applicable to game design, to stay on subject, our literature selection only admits works that specifically target game content creation. An initial endeavor in mixed-initiative targeted at generating game bits is the work of Liapis et al. \cite{Liapis2012, Liapis2012a}. They introduced an approach for generating personalized content based on user preference regarding visual aesthetics. The framework presents eight spaceships that were optimized offline, from which the user selects their preferred one. The system then generates a new set of spaceships, enhancing the visual features behind the user’s choice via an adaptive aesthetic model. After a few iterations, the newly evolved content is more likely to be appealing to the user. 

We have chosen more recent literature to include in this class of game content. We describe two works below, which will be subject to further analysis using the framework proposed in section~\ref{MIclassification}.

\subsubsection{AI Spaceship Generator}

\begin{figure}[h]
  \centering
  \includegraphics[width=0.7\linewidth]{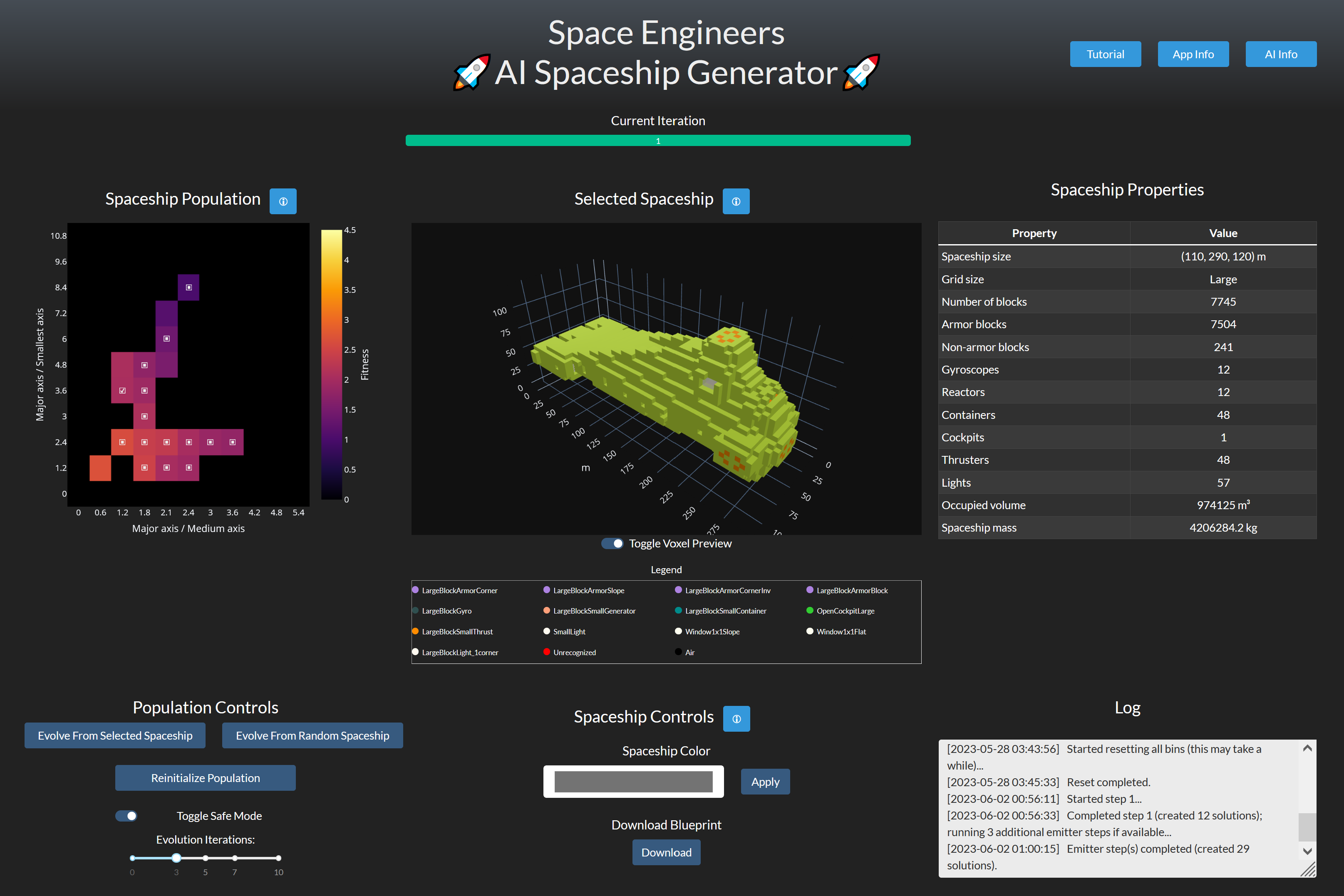}
\caption{Interface of the AI Spaceship Generator \cite{Gallotta2023}. Top left --- the spaceship population; top middle --- 3D preview of the selected spaceship; top right --- a list of the spaceship properties; bottom left --- controls to generate more spaceships or reset the population; bottom middle - option to change the spaceship color and option to download its blueprint; bottom right - application logs. (Available at \url{https://github.com/arayabrain/space-engineers-ai-spaceship-generator})}
\label{figGallotta2023}
\end{figure}

Gallotta et al. \cite{Gallotta2023} presented an approach for content generation aligned with the user's preferences by incorporating preference learning into quality-diversity algorithms. To showcase this approach, they implemented an application to generate 3D spaceships for Space Engineers, a game that allows players to construct block-based structures. The interface of the AI Spaceship Generator (Fig.~\ref{figGallotta2023}) presents a grid with an initial population of spaceship suggestions. The user can select any of these suggestions and see its 3D preview along with a list of its in-game properties. If the user appreciates the selected spaceship, they can press the ``Evolve from Selected Spaceship'' button and the system will generate more suggestions via a predictive model of the preferences of the human designer. Otherwise, the user can press the ``Evolve from Random Spaceship'' button to be presented with more random suggestions. As the spaceship population is updated, the user can select a new spaceship and repeat the process. With each generation based on user selection, the system continues to learn preferences for the newly evolved content to stay aligned with the user’s intent. The user can also change the color of the spaceship and a few generation settings. The spaceship resulting from the iterative process can be downloaded and imported into Space Engineers.
Extending procedural content generation systems to mixed-initiative settings by solely accounting for user selection is a common but limited approach if the aim is to afford human-computer co-creativity. We discuss this further in Subsection~\ref{LiteratureClassification}.

\subsubsection{Sprite Editor for Pixel Art Characters}

\begin{figure}[h]
  \centering
  \includegraphics[width=0.6\linewidth]{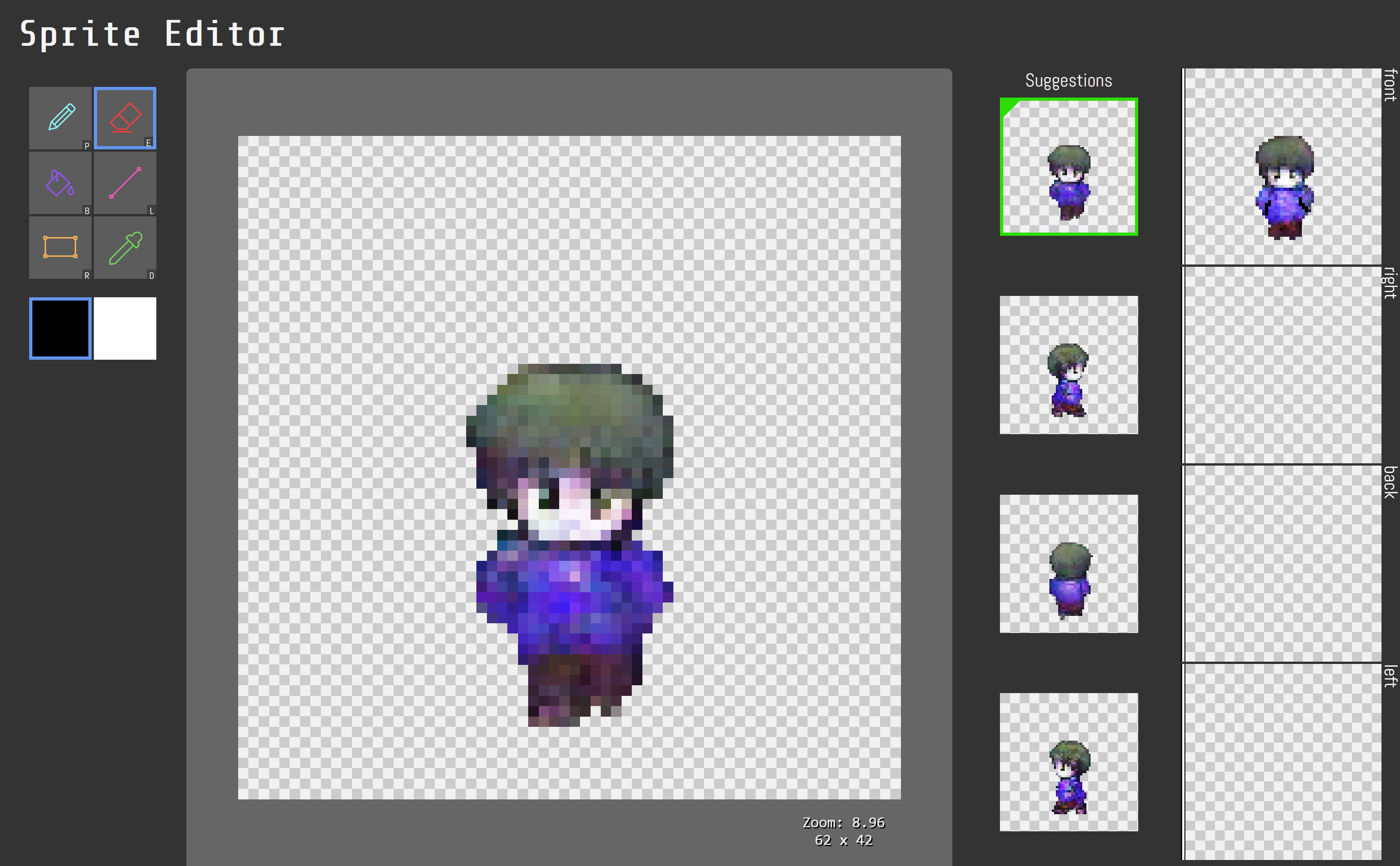}
\caption{Interface of the sprite editor \cite{Coutinho2024}. Left --- the toolbar; middle --- the main canvas displaying the selected pose view; right --- a panel with the four pose views and a panel with the suggestion slots for each pose view on its left side. (Available at \url{https://fegemo.github.io/sprite-editor-gm/})}
\label{figCoutinho2024}
\end{figure}

Coutinho and Chaimowicz \cite{Coutinho2024} introduced an approach for creating sprites for pixel art characters facing four directions (front, back, left, and right). To this end, they developed a sprite editor  (Fig.~\ref{figCoutinho2024}) in which the human agent can interact with the computational agent to generate the various poses. The system contains a main canvas for the user to view and edit their selected pose. The human agent initiates the collaborative process by starting to draw the character on the main canvas. At any time, the designer can drag the pose they drew into a suggestion slot for a different pose (e.g., dragging the front view into the right view suggestion slot). A trained generative model takes this image as input and generates an image for the target pose, displaying it in the suggestion slot. The human agent can engage in multiple iterative steps, either directly editing any pose view or asking the computational agent to generate a new suggestion from any view that is not the same direction as the target.

\subsection{Game Space}\label{gameSpace}

Game space is the main field of application of mixed-initiative approaches to game design. A fair sample of current works is concerned with the level design domain, in which the game space creation is usually done through the positioning and rearrangement of tiles or components to form a map in a way that results in a valid playable level.  Some of the tools we analyzed within this domain include Tanagra \cite{Smith2010,Smith2011}, Ropossum \cite{Shaker2013,Shaker2013a}, the progression design tool for the Refraction game \cite{Butler2013}, Ludoscope \cite{Karavolos2015},  Evolutionary Dungeon Designer \cite{Baldwin2017,Baldwin2017a,Alvarez2018,Alvarez2018a,Alvarez2019}, Mixed Initiative Procedural Dungeon Designer \cite{Walton2020}, Morai Maker \cite{Guzdial2019}, Lode Encoder \cite{Bhaumik2021}, and Baba is Y’all \cite{Charity2020,Charity2022}. The level representation (e.g., dungeons, platforms) for each of these tools depends on their targeted game genre. We have selected three works for in-depth analysis, which we present below.

\subsubsection{Sentient Sketchbook} 
\begin{figure}[h]
  \centering
  \includegraphics[width=.6\linewidth]{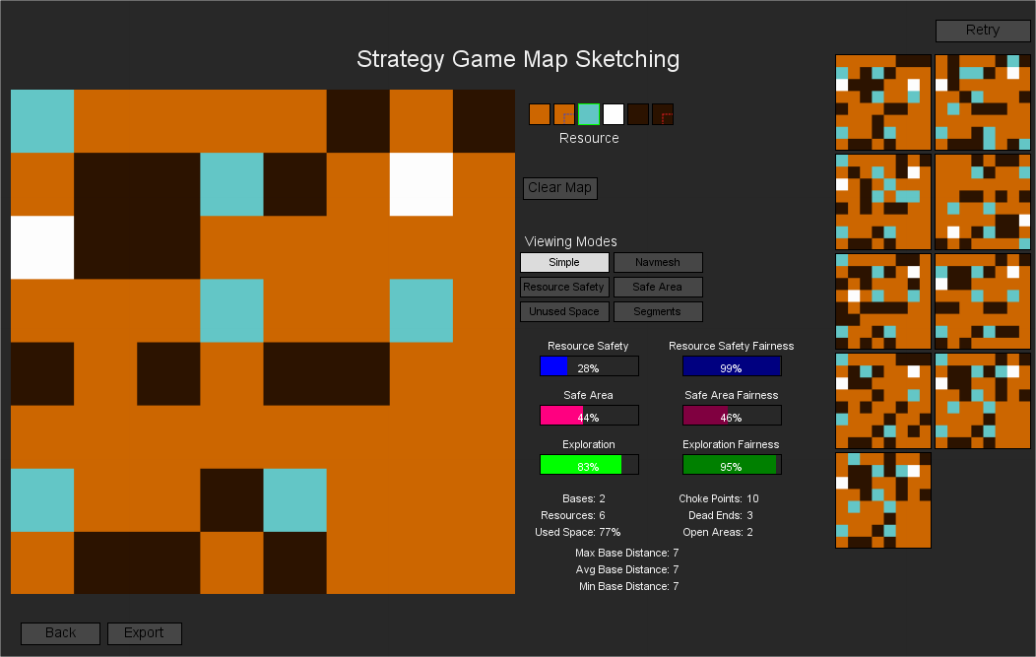}
\caption{The Sentient Sketchbook tool \cite{Yannakakis2014}. Left --- the sketch editor; middle --- the tile palette, the map display menu, and an overview of the map’s fitness dimensions and metrics; right --- the
automatically generated map suggestions. (Available at \url{https://www.sentientsketchbook.com/download.php})}
\label{figYannakakis2014}
\end{figure}

Liapis et al. \cite{Liapis2013,Yannakakis2014} created a tool for the design of game levels. The Sentient Sketchbook (Fig.~\ref{figYannakakis2014}) enables users to create and edit levels using map sketches as simplified representations of levels. As the human designer engages with the tool through sketching, the system dynamically provides suggestions that are evolved variations of their current sketch. These suggested alternative map designs are generated using a genetic search algorithm that aims to optimize the levels’ fitness scores or produce visually diverse maps. The tools’ interface displays the fitness scores and other metrics of the current sketch for the user to compare them to the scores and metrics of each suggestion. The human agent can choose to replace their current sketch with a suggested alternative at any time. Once satisfied with the outcome, the designer can request the system to convert the map sketch into a playable 2D or 3D level with enhanced detail.

\subsubsection{RL Brush} 
\begin{figure}[h]
  \centering
  \includegraphics[width=.6\linewidth]{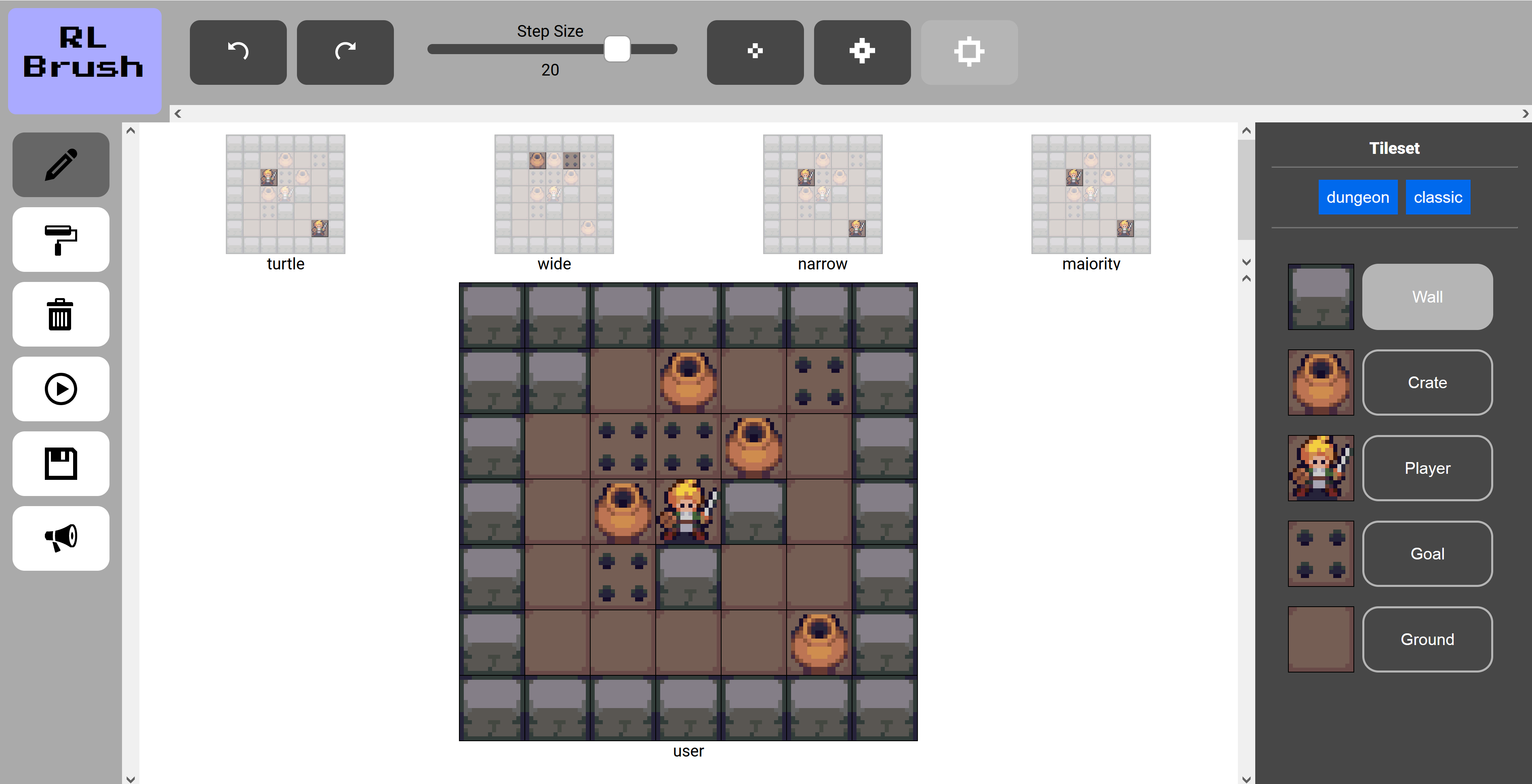}
\caption{The RL Brush tool \cite{Delarosa2021}. Top --- the \emph{step} and \emph{tool radius} parameters; left --- the toolbar to edit, erase, delete, play, and save the current level; middle --- the editable grid of the current level and the four AI suggestions on top; right - the tileset. (Available at \url{https://rlbrush.app/})}
\label{figDelarosa2021}
\end{figure}

Delarosa et al. \cite{Delarosa2021} presented a mixed-initiative tool for designing tile-based levels, with a version specifically tailored for the puzzle game Sokoban. RL Brush (Fig.~\ref{figDelarosa2021}) uses a set of AI models to interactively suggest incremental modifications while the human designer is editing the level. The system presents four suggestions at a time generated by reinforcement-learning-based agents: a \emph{narrow} agent that can change a specified tile at each step, a \emph{turtle} agent that can move to a next tile in the map grid and modify passed tiles, a \emph{wide} agent that has control to edit any tile in the map, and a \emph{majority} agent that suggests tile mutations if the majority of the other three agents have suggested that same mutation. The display of new suggestions is triggered by user interactions only. The user can accept any of these suggestions or override them by manually editing the grid for the current level. The tool also allows the user to fine-tune the performance of the AI models by providing two hyperparameters for the user to control. The \emph{step} parameter defines the number of interactions for the manager of the models to recursively call itself, so that a higher step value allows for more than one modification to the map. The \emph{tool radius} parameter controls how big is the window of tiles that the artificial agents can view and modify. By generating modifications one step at a time, this approach results in a creative process in which participants gradually build the level tile by tile.

\subsubsection{miWFC} 
\begin{figure}[h]
  \centering
  \includegraphics[width=.6\linewidth]{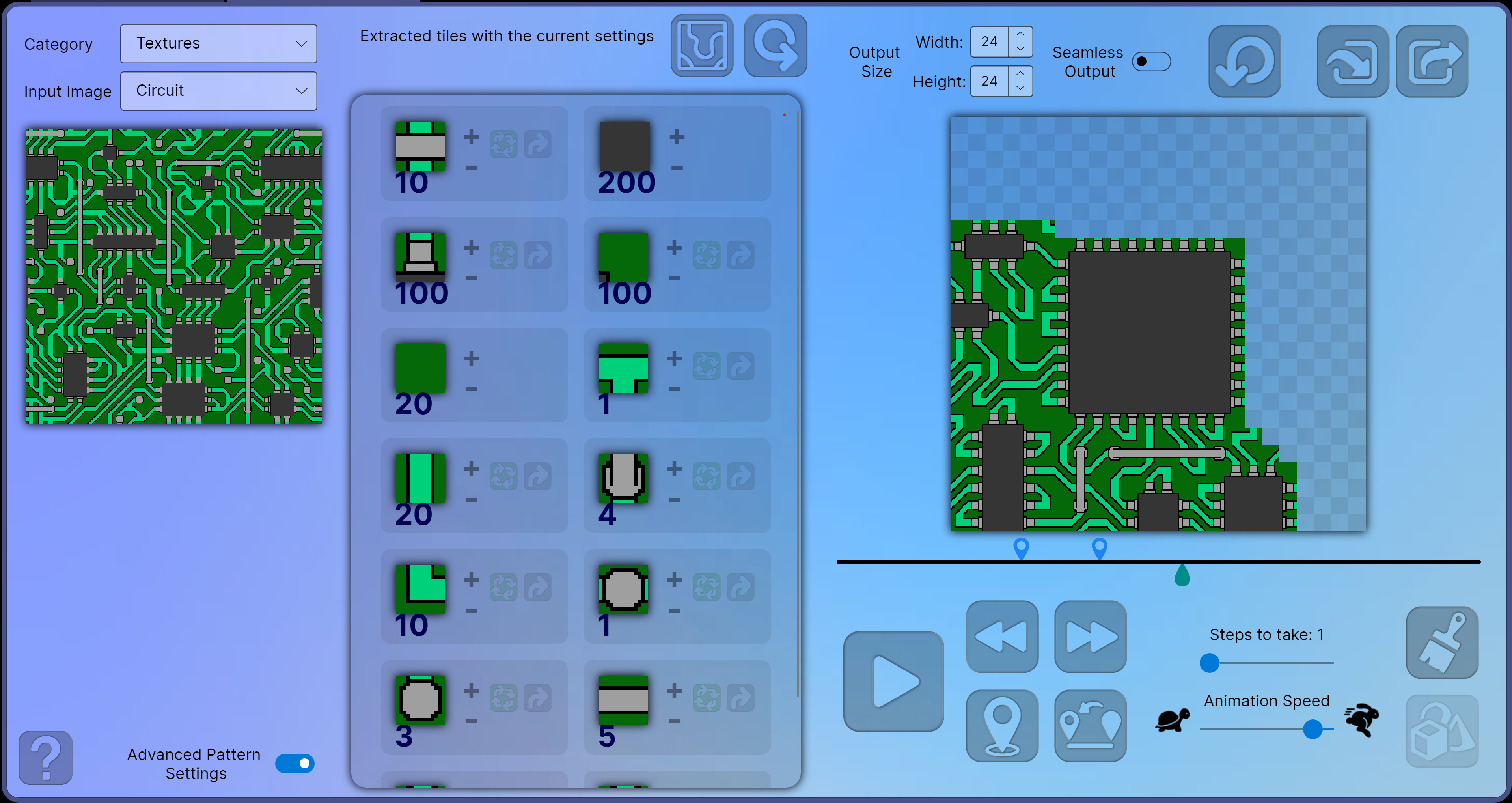}
\caption{The miWFC tool \cite{Langendam2022}. Left --- a list with several example input images with the selected one displayed; middle - the panel for manipulating tile properties; right --- the current image accompanied by a timeline indicating the progress of the generation and a number of controls for adding markers and navigating through the timeline. The screen for direct manual editing is accessible from the brush button in the bottom right corner. (Available at \url{https://github.com/ThijmenL98/miWFC})}
\label{figLangendam2022}
\end{figure}

Langendam and Bidarra \cite{Langendam2022} introduced a mixed-initiative approach to Wave Function Collapse, which can be used to create various types of maps or other assets such as textures and 2D objects. The miWFC tool (Fig.~\ref{figLangendam2022}) incorporates several interactive features that grant greater control to the human designer over the content generated by the system. While the system displays the gradual generation of the image through an animation, the tool enables the user to navigate through the timeline of the generative process. At any point, the user can add markers to register the current progress and easily backtrack to any of these markers if desired. Additionally, they have the option to advance or reverse step by step through the generation. The human agent can also directly manipulate the tiles of the image at any stage of the generative process. Several tools are available for them to add desired tiles or select areas of the map to erase or preserve. Finally, the tool allows the human agent to control several properties of the tiles prior to generation, which can significantly influence the output of the computational agent. These properties include the ability to exclude specific tiles, set weights to prioritize the appearance of certain tiles over others, determine whether tiles can be rotated and/or mirrored, and even manipulate the distribution of tile weights across the output space using heatmaps.

\subsection{Game Behaviors} 
To the best of our knowledge, there is a dearth of mixed-initiative approaches specifically targeting game behaviors, such as the mechanics of game elements. Therefore, we present here a single tool that we have identified as fitting into this category.

\subsubsection{Pitako} 
\begin{figure}[h]
  \centering
  \includegraphics[width=.6\linewidth]{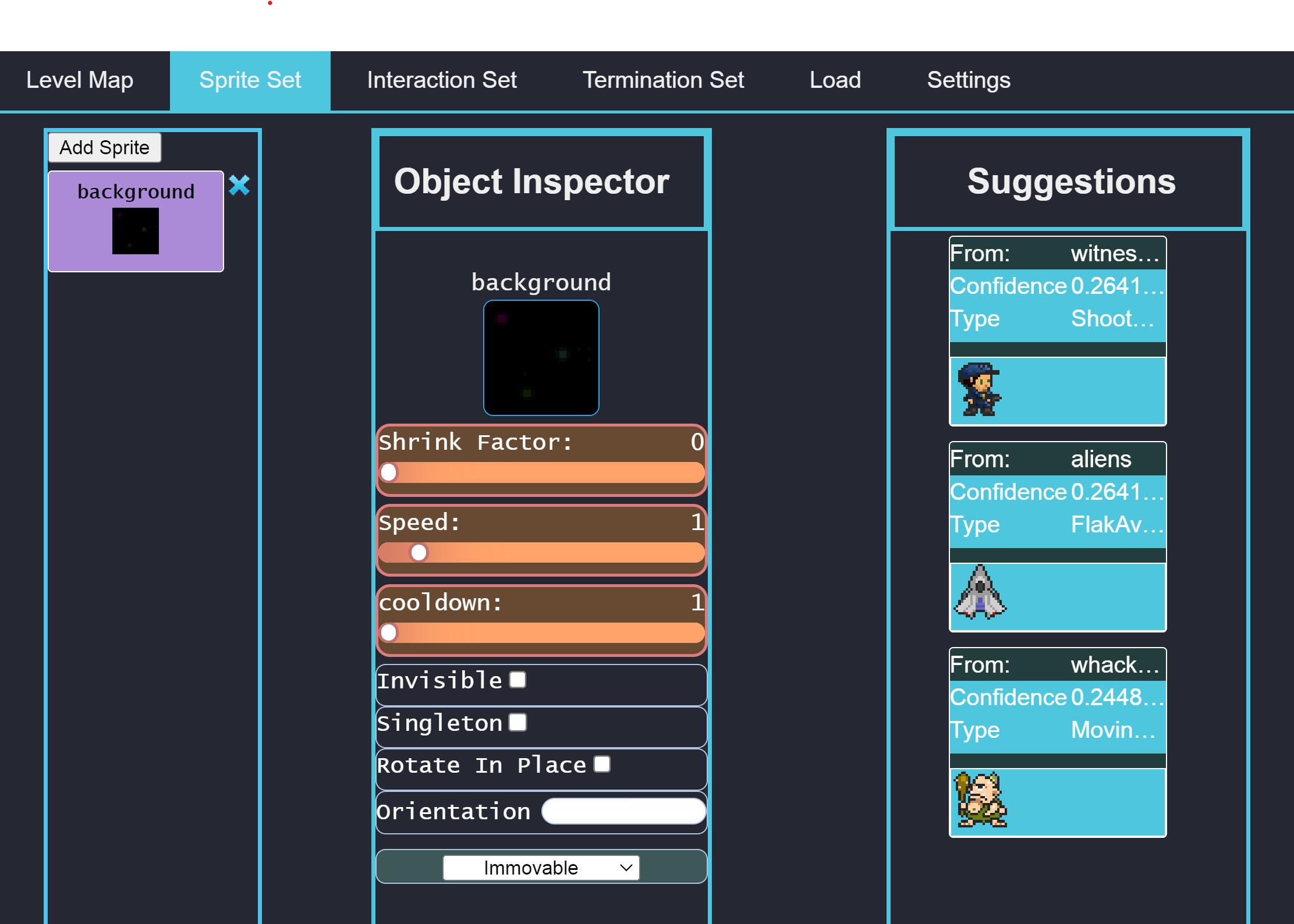}
\caption{Sprite Recommendation in the Pitako tool \cite{Machado2019}. Top --- a menu featuring tabs for the several tasks within the game design process; left --- the game's sprite set, initially containing only a default background sprite; center --- the sprite property editor, enabling the selection of an image and the configuration of its behavior; right --- a list of the system-generated suggestions. (Available at \url{https://github.com/tiago-lam/Recommender-System-for-Game-Design})}
\label{figMachado2019}
\end{figure}

Machado et al. \cite{Machado2019, Machado2019a} developed a tool for designing tile-based games that applies the concept of a recommender system to suggest mechanics and dynamics learned from existing games. The system is built on Video Game Description Language (VGDL), which considers sprites as objects with a graphical representation and an associated behavior. The creative process can begin by adding sprites to the game in development. For instance, the user can define a sprite behavior as a ShootAvatar and then attach the image of the hero to it. Having a catalog of a vast number of games, the system can provide suggestions of sprites to the designer, who is free to accept them or not (Fig.~\ref{figMachado2019}). If the user accepts a recommended sprite, it is imported to the game sprite set along with any others it may carry (e.g., picking a sprite that can shoot will also import the sprite for the projectile behavior). In addition, the system also suggests interactions for pairs of sprites (e.g., enemy sprite kills avatar sprite), taking its cues from the existing games in the catalog. If the user selects a recommendation, the interaction is automatically implemented in the game.

This mixed-initiative approach primarily targets content creation within the class of game behaviors. However, it also extends to the game space and game design classes by enabling the placement of sprites on the level grid and the definition of game termination rules. Still, it only serves as a support tool for these two classes as the computational agent lacks initiative. 

\subsection{Game Systems} 

The literature review has revealed a gap in mixed-initiative works focused on the creation of game systems. Most approaches are merely generative, lacking an iterative process between humans and computers. Such can be attributed to the nature of this type of game content, which often aims to simulate real-life interactions (e.g., the behavior of an army in battle) and therefore requires little creativity. Thus, the mere generation of game systems may be satisfactory enough for designers. It is also a demanding task that involves programming complex interactions among multiple entities. Furthermore, it requires the in-game simulation of these interactions for assessment and refinement, which can be highly time-consuming and significantly prolong each iteration. We were only able to find one work in the scope of game systems that we could classify as a mixed-initiative approach. 

\subsubsection{EvolvingBehavior} 
\begin{figure}[h]
  \centering
  \includegraphics[width=.8\linewidth]{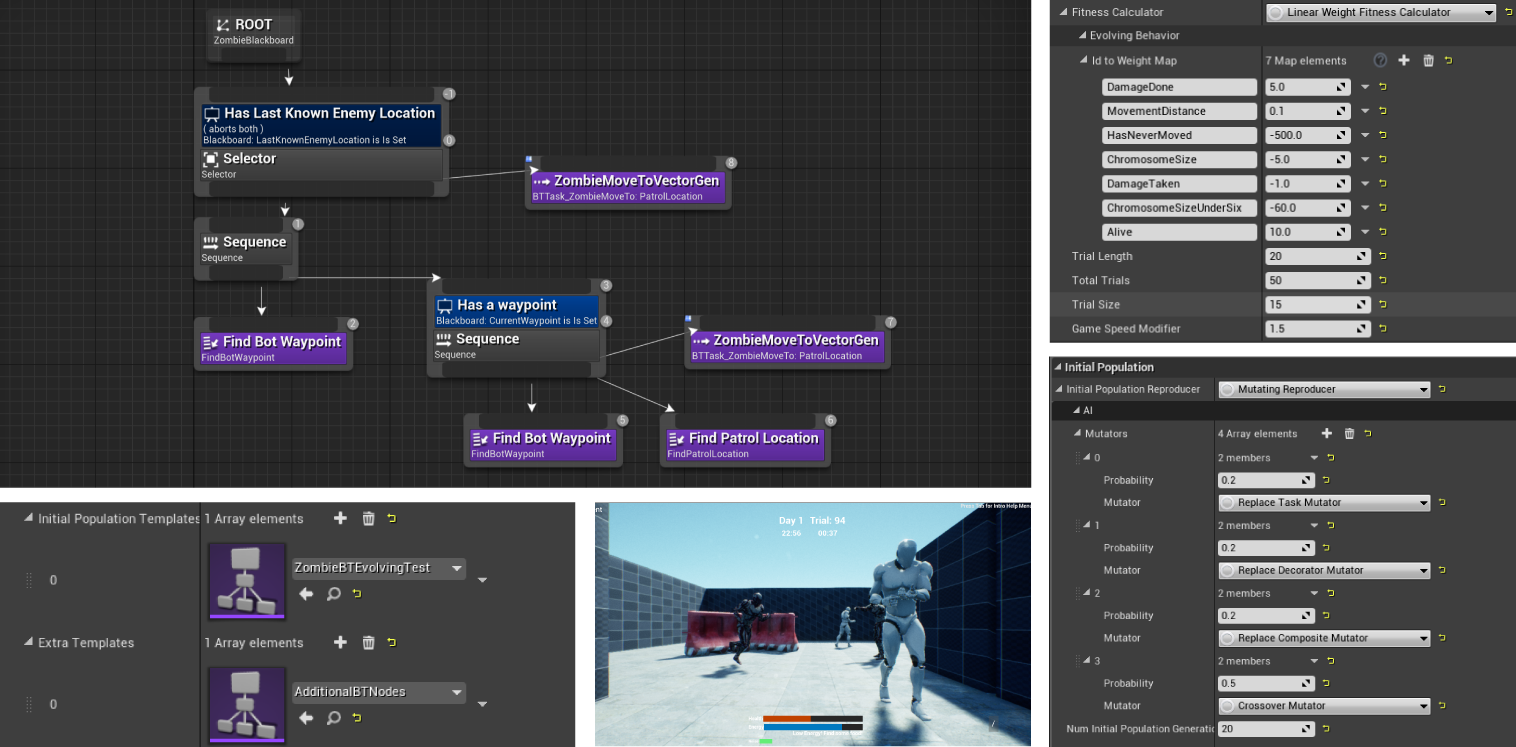}
\caption{Some screens of the EvolvingBehavior tool \cite{Partlan2022}. Top left --- an evolved tree; bottom left --- template collection for starter behavior trees; bottom center --- visualization of the NPCs' behavior; top right --- settings to define fitness functions; bottom right --- mutation properties for the initial population. (Available at \url{https://evolvingbehavior.npc.codes/}) }
\label{figPartlan2022}
\end{figure}

Partlan et al. \cite{Partlan2022} introduced a tool for crafting non-player character (NPC) behavior by evolving behavior trees in Unreal Engine 4 (Fig.~\ref{figPartlan2022}). The process begins with the human designer providing an initial draft of a tree and other materials, such as a library of pre-defined nodes and templates for generating additional nodes. The system can then modify and generate behavior trees, even if the initial tree is empty or incomplete. In order for the evolution of behavior trees to meet the goals of the human designer, they must define a fitness function by setting fitness measurement keys. The user also needs to define several parameters of the evolutionary process. Given that the evolved trees can be manually edited, in practice, the human designer can modify them and use them as input for a new round of evolution, engaging in an iterative process. However, this work was not selected for further analysis in section~\ref{MIclassification}. This is because it is currently in a state where user interaction with the tool is too complex for a regular designer without coding skills and expert knowledge of the system to easily work with it. As acknowledged by the authors, it is challenging to define the fitness function and adjust all the parameters without prior knowledge. Additionally, the generation and subsequent simulation of behavior trees are time-consuming. The delay would likely discourage designers from participating in a new evolutionary process.

\subsection{Game Scenarios}

Narrative is a type of content within the game scenarios class that is highly popular in co-creative tools outside the domain of video games. While not their specific focus, most of these tools can be easily applied to game design, particularly for story-driven games. Large language models like GPT-3 \cite{Brown2020} can be used for the purpose of generating stories.  An example of this is Wordcraft \cite{Yuan2022}, a text editor for co-creative story writing with the assistance of a large language model. However, co-creative storytelling does not necessarily entail both agents contributing to the narrative in prose form. In Writing Buddy \cite{Samuel2016}, the computational agent offers suggestions of character actions that are consistent with the narrative. This tool offers a playful setting for co-creative writing by establishing narrative goals for the user to complete. Similarly, Loose Ends \cite{Kreminski2022} provides the user with narrative goals that involve a structured sequence of plot events and can propose new goals consistent with the evolving story throughout the creative process.

In some cases, narrative design is not solely conducted during game development but also involves in-game content creation. This occurs when narrative creation is either the very concept of the game or is integrated into its dynamics, such as in dialogues between the player and non-player characters. Below, we present two distinct MI-CCy approaches for game scenarios: one that applies the creation of narrative structures to a specific game genre and another in which storytelling is an integral part of the game itself.

\subsubsection{Story Designer} 
\begin{figure}[h]
  \centering
  \includegraphics[width=.6\linewidth]{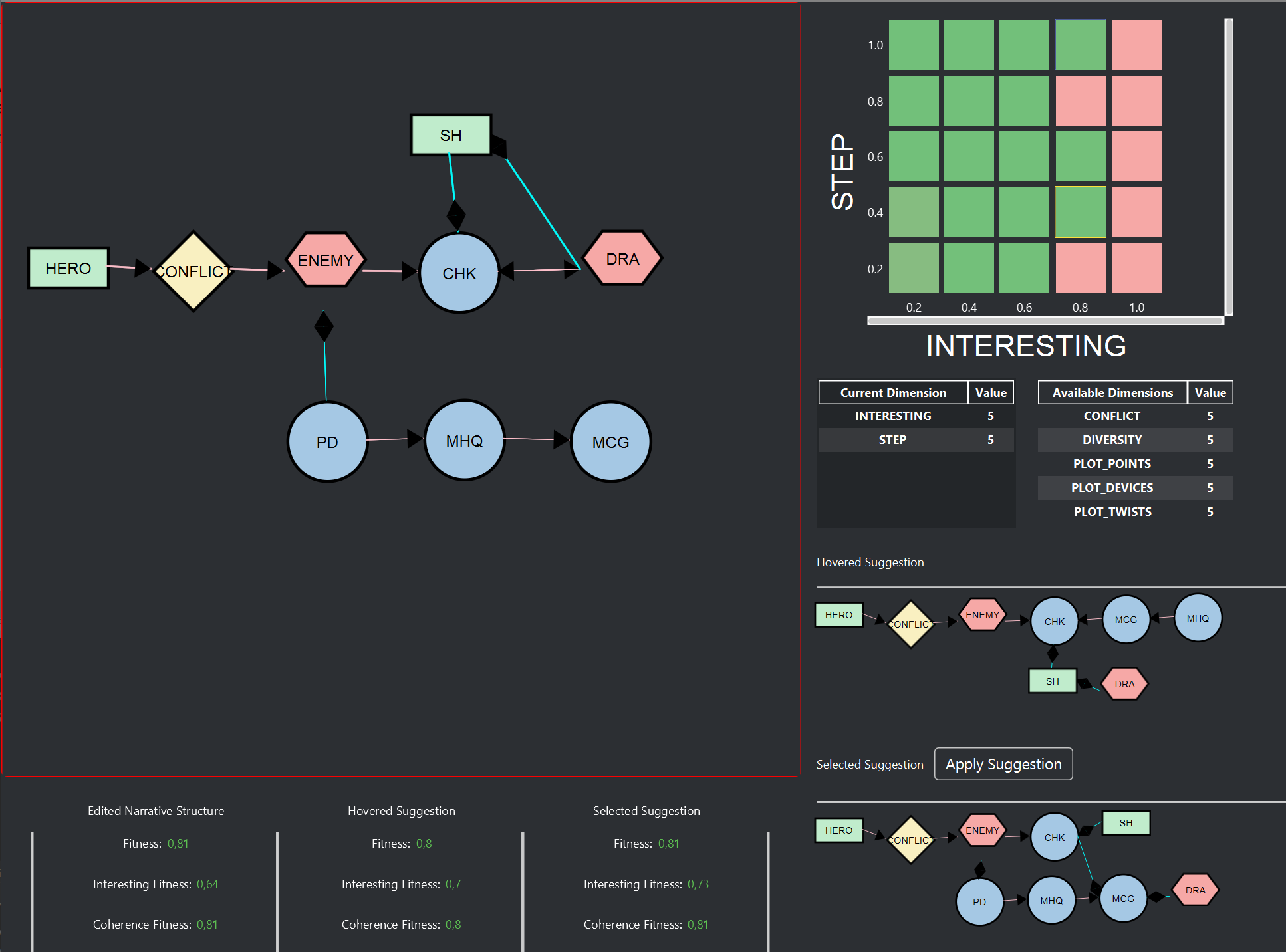}
\caption{The Story Designer tool \cite{Alvarez2022}. Center --- the editor for the main narrative graph; right --- the suggestion grid, the list of dimensions that the user can choose for the grid, the suggestion they are hovering, and the selected suggestion; bottom --- display of fitness metrics of the main graph and of the suggestions. (Available at \url{https://github.com/mau-games/eddy}) }
\label{figAlvarez2022}
\end{figure}

The Evolutionary Dungeon Designer (EDD) is a mixed-initiative tool for designing 2D dungeons and the rooms that compose them. EDD is part of an ongoing research project and has been progressively expanded over several works \cite{Baldwin2017,Baldwin2017a,Alvarez2018,Alvarez2018a,Alvarez2019,Alvarez2021}. As a next step, Alvarez et al. \cite{Alvarez2022} built an approach to design narrative structures integrated in EDD. Story Designer (Fig.~\ref{figAlvarez2022}) uses tropes in the form of nodes that can be interconnected in a graph to construct a complete narrative structure. The creative process begins with the system providing a basic sample graph HERO → CONFLICT → ENEMY. Users can manually edit the narrative graph by adding and connecting nodes representing characters, their roles, goals, and events. This triggers an evolutionary algorithm that generates alternative suggestions of narrative structure in real time that the user can inspect and choose to replace the current narrative design. The algorithm is an adaptation of MAP-Elites, which evaluates the set of solutions according to metrics such as interestingness and coherence. The system displays the suggestions in a matrix arranged by their values across specified dimensions of interest. Solutions with optimal fitness are shown in dark green, while dark red cells mean the best solution for those values is still not feasible. The human designer can visualize two dimensions at a time and switch between them as needed. As Story Designer is linked to the level design tool of EDD, the suggested narrative graphs are constrained by the existing content in the level’s map. 

\subsubsection{1001 Nights} 
\begin{figure}[h]
  \centering
  \includegraphics[width=.6\linewidth]{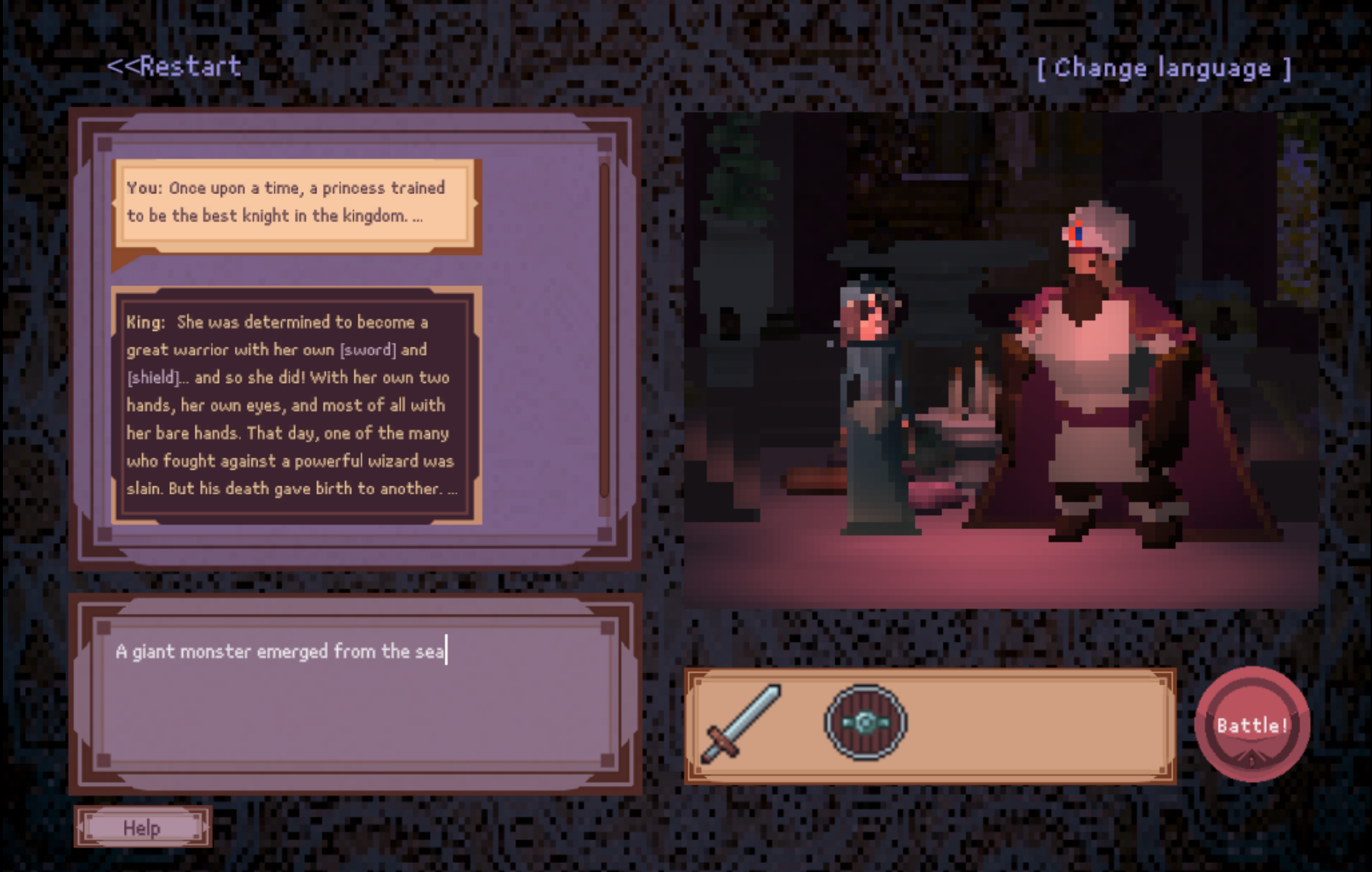}
\caption{The 1001 Nights game \cite{Sun2022}. Left --- a dialogue interface containing text blocks from the story told by the player and the King, along with a text box at the bottom for the player to write; right --- a screen featuring the game characters, in which the player can collect the generated items and add them to the inventory at the bottom; bottom right corner: the button to initiate the battle. (Available at \url{https://cheesetalk.itch.io/one-thousand-and-one-night})}
\label{figSun2022}
\end{figure}

Sun et al. \cite{Sun2022} developed 1001 Nights, a co-creative storytelling game that transforms story elements into game mechanics. The game features two characters: Shahrzad, controlled by the player, and the King, controlled by an AI model. The player needs to tell a story by typing it into a chat-like interface (Fig.~\ref{figSun2022}), and the King responds by continuing the tale. Consequently, the narrative unfolds through alternating turns. If the King mentions items such as ``sword'', ``dagger'', and ``shield'' in the story, the player can convert these words into weapons within the game to use in a battle against the King. Thus, the goal is for the player to narrate a story that prompts the system to include these words. This goal serves to constrain the content generated by both the human and the system, ensuring the story remains on topic. To encourage the player to continue developing the narrative, the game requires them to possess at least two attack weapons to be able to defeat the King. The words the game allows to convert into equipment are consistent with the ancient context of the classic One Thousand and One Nights folktales.

\subsection{Game Design}
There are few MI-CCy approaches that address high-level concepts, such as the rules and objectives of a game. Most existing tools are instead tailored for a specific game or game genre. So far, when the creative task involves the ideation of a novel game, it is still uncommon to grant decision-making authority to the computational agent, relegating them to a passive role in the process. Still, there are some works that deviate from this pattern. Among the tools we have reviewed, CADI and Gamika fall into this category but were not selected. CADI \cite{Mobramaein2018} is a tool that employs a conversational interface for agents to design variations of the game Pong. Gamika \cite{Nelson2017} is a system for creating 2D physics-based games for mobile devices, which parametrizes the design space into 284 features that the user can manipulate.

\subsubsection{Germinate} Kreminski et al. \cite{Kreminski2020a} introduced a mixed-initiative casual creator for designing rhetorical games. This framework builds upon Gemini, an abstract game generator, but has a more user-friendly interface tailored for casual users. Through the browser-based interface, human agents can specify certain properties they wish to see in the generated games. These properties are defined using four types of cards:  entity cards, which represent game objects with a graphical representation (emoji) that react to interactions with the player or other entities; resource cards, which describe quantitative values linked to the objectives of the game; relationship cards, which establish connections between entities and/or resources; and trigger cards, which describe the outcomes of specific trigger events. Gemini then generates a batch of games based on these specified properties, which are displayed in the interface of the tool. Each game includes a set of rules, and respective behaviors and interactions that align with the user's expressed intentions.

\begin{figure}[h]
  \centering
  \includegraphics[width=.6\linewidth]{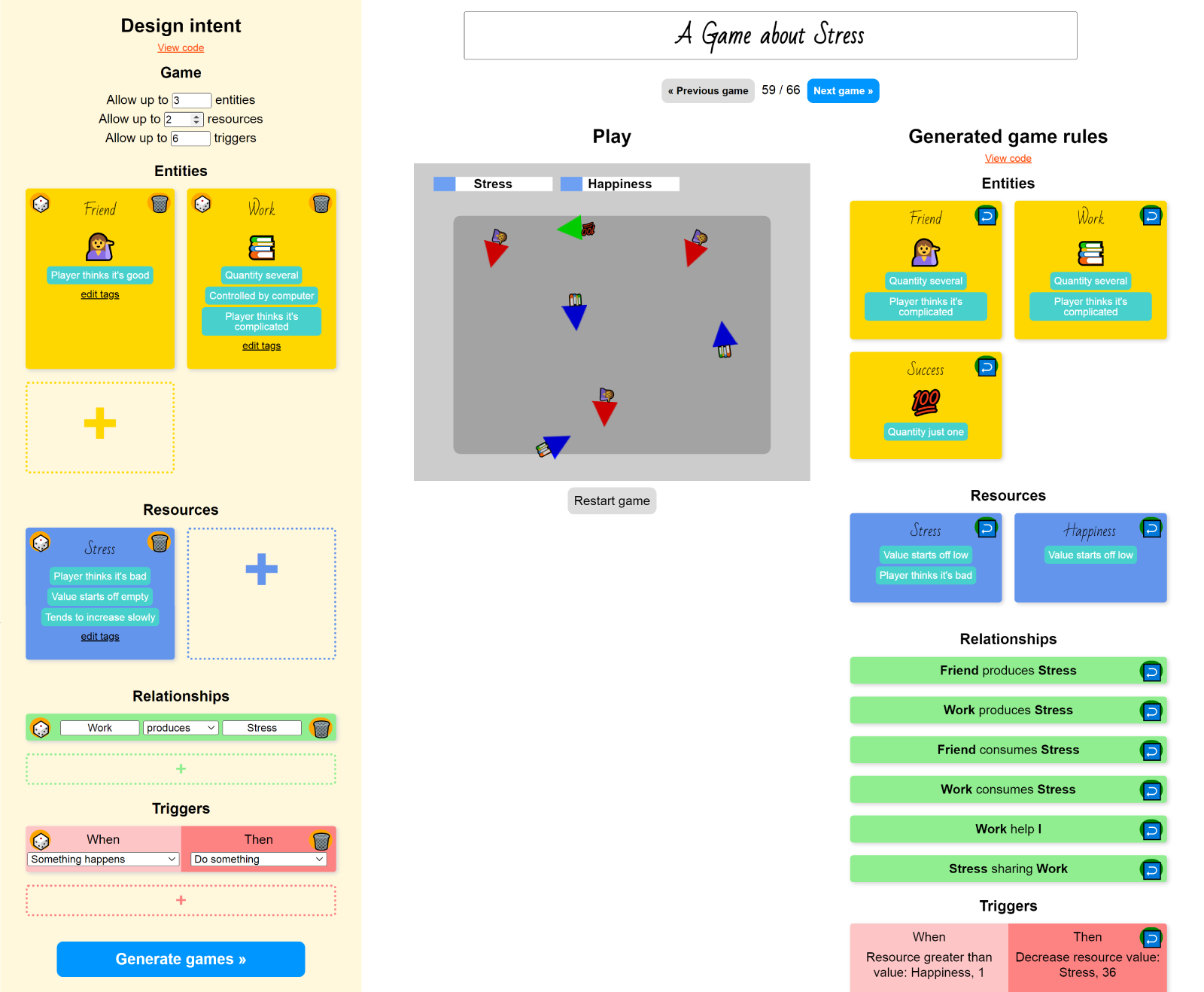}
\caption{The Germinate tool \cite{Kreminski2020a}. Left --- properties specified by the user; center --- a playable version of one of the generated games (a single game is shown at a time); right --- the properties of the generated game. (Available at \url{https://github.com/ExpressiveIntelligence/Germinate})}
\label{figKreminski2020a}
\end{figure}

For instance, a user interested in creating a stress-themed game might define entities such as ``Friend'' and ``Work'', a resource called ``Stress'', and the relationships and triggers they want to showcase. This example is illustrated in Fig.~\ref{figKreminski2020a}.
The user can explore the several generated games to assess if there are any interesting results and then make adjustments to the properties to better reflect their intent, allowing for iterative refinement of the game. 
This approach is focused on starting from high-level rhetorical goals and then working on details to meet those goals. To articulate the conceptual essence of the game, Germinate also relies on the creation of a game system and game behaviors.

\subsection{Overview}

Each work presented in this section falls into one of the six classes of game content we previously defined. However, these tools may tackle more than one class. Table~\ref{workGameContentClasses} provides a list of the selected works, with all covered classes of game content explicitly marked. 

\begin{table}
\caption{Categorization of Works by the Tackled Classes of Game Content}
\label{workGameContentClasses}
\centering
\begin{tabular}{lcccccc} 
\toprule
Work & \begin{tabular}[c]{@{}c@{}}Game\\Bits\end{tabular} & \begin{tabular}[c]{@{}c@{}}Game\\Space\end{tabular} & \begin{tabular}[c]{@{}c@{}}Game\\Behaviors\end{tabular} & \begin{tabular}[c]{@{}c@{}}Game\\Systems\end{tabular} & \begin{tabular}[c]{@{}c@{}}Game\\Scenarios\end{tabular} & \begin{tabular}[c]{@{}c@{}}Game\\Designs\end{tabular} \\ 
\midrule[0.7pt]
\begin{tabular}[c]{@{}l@{}}\textit{AI Spaceship Generator}\\Gallotta et al. (2023)\end{tabular} & x &  &  &  &  &  \\
\midrule[0.2pt]
\begin{tabular}[c]{@{}l@{}}\textit{Sprite Editor for Pixel Art Characters}\\Coutinho and Chaimowicz (2024)\end{tabular} & x &  &  &  &  &  \\
\midrule[0.2pt]
\begin{tabular}[c]{@{}l@{}}\textit{Sentient Sketchbook}\\Liapis et al. (2013)\end{tabular} &  & x &  &  &  &  \\
\midrule[0.2pt]
\begin{tabular}[c]{@{}l@{}}\textit{RL Brush}\\Delarosa et al. (2021)\end{tabular} &  & x &  &  &  &  \\
\midrule[0.2pt]
\begin{tabular}[c]{@{}l@{}}\textit{miWFC}\\Langendam and Bidarra (2022)\end{tabular} & x & x &  &  &  &  \\
\midrule[0.2pt]
\begin{tabular}[c]{@{}l@{}}\textit{Pitako}\\Machado et al. (2019)\end{tabular} &  & x & x &  &  & x \\
\midrule[0.2pt]
\begin{tabular}[c]{@{}l@{}}\textit{EvolvingBehavior}\\Partlan et al. (2022)\end{tabular} &  &  & x & x &  &  \\
\midrule[0.2pt]
\begin{tabular}[c]{@{}l@{}}\textit{Story Designer}\\Alvarez et al. (2022)\end{tabular} &  &  &  &  & x &  \\
\midrule[0.2pt]
\begin{tabular}[c]{@{}l@{}}\textit{1001 Nights}\\Sun et al. (2022)\end{tabular} &  &  &  &  & x &  \\
\midrule[0.2pt]
\begin{tabular}[c]{@{}l@{}}\textit{Germinate}\\Kreminski et al. (2020)\end{tabular} &  &  & x & x &  & x \\
\bottomrule
\end{tabular}
\end{table}

\section{Mixed-initiative Classification}\label{MIclassification}

Getting started with mixed-initiative approaches often requires acquaintance with the various existing systems and identifying what distinguishes them, what techniques are used, what is novel about them, how is mixed-initiative present, what role the human and the computer play, among other questions. Surveys and overviews in the literature serve as valuable resources for providing a broad introduction to the subject. In particular, Lai et al.\cite{Lai2022} provide an overview of the techniques commonly used in mixed-initiative systems for game content creation and its inherent challenges. While researchers and practitioners undoubtedly benefit from this knowledge, we emphasize that individuals implementing a MI-CCy system should, foremost, reflect on what kind of human-computer interaction they want to convey and then choose the technique that will better deliver that intent. To do so, it is also notably useful to have tools that allow us to carry out our own analysis when designing a mixed-initiative system so we can reflect on its novelty and contributions to human-computer co-creativity.

Deterding et al. \cite{Deterding2017} emphasize how the human and the computer iteratively respond to each other through a variety of possible actions. Following this idea, Spoto and Oleynik \cite{Spoto2017} introduced a framework that allows the comparison of mixed-initiative systems regarding the flow of the creative process. The flow is visualized through a graph that considers seven subprocesses of the creative process, namely ideate, constrain, produce, suggest, select, assess, and adapt, which can be executed by both the human and the computer. The graph illustrates the sequence of actions through directional links, usually outlining internal interactive loops. Muller et al. \cite{Muller2020} expand on this framework, addressing mixed-initiative scenarios with generative AI systems. Grabe et al. \cite{Grabe2022} also adapt the framework to co-creativity with Generative Adversarial Networks (GANs), identifying four primary patterns of interaction between humans and GANs. Focused on interaction design in co-creative systems, Rezwana and Maher \cite{Rezwana2021,Rezwana2022} presented COFI, a framework that describes several interaction components to represent the space of possibilities for interaction between humans and AI and with the shared creative content. We consider these frameworks to be valuable, making it possible to distinguish between approaches in terms of how the creative process admits the initiative of both agents involved and how they interact with each other. However, they make little to no mention of the degree of mixed-initiative, the real impact each agent has on the creative process, and how much their collaboration is facilitated.

\subsection{MI-CCy Quantifier: A Framework for Mixed-Initiative Co-Creativity Quantification}

Liapis et al. \cite{Liapis2016a} consider it ingenuous to expect human and computational agents to be equal partners in the creative process. We admit that participants involved in co-creation cannot be exactly equal throughout the various stages of the process, which is also true for human-human co-creativity, as each person has their own personality, knowledge, capabilities, and past experience. Nonetheless, we argue that such should be explored as one of the possible settings for future mixed-initiative tools. Liapis et al. \cite{Liapis2016a} reiterate that current approaches favor human control, with the user having the final say on the decisions taken. We hope that new approaches can challenge this view. As mentioned earlier, a high degree of human initiative combined with a low degree of computational initiative prevents the designer from fully leveraging the unique capabilities of AI systems. Conversely, a very high degree of computational initiative constrains the designer's creativity and control. It is our belief that working as colleagues in a process in which collaboration is continuous and facilitated allows the human and the machine to learn from each other productively and can maximize the capabilities of both.
This perspective stems from the observation that the most fruitful co-creative interactions often occur when both agents are able to take initiative, influence the direction of the creative outcome, and respond meaningfully to each other's contributions. By allowing humans and AI to work as equals in the creative process, we can foster a richer and more dynamic collaboration that goes beyond the restrictive roles of mere assistance or automation. This enables more creative exploration and the emergence of surprising outcomes.

\begin{figure}[h]
  \centering
  \includegraphics[width=\linewidth]{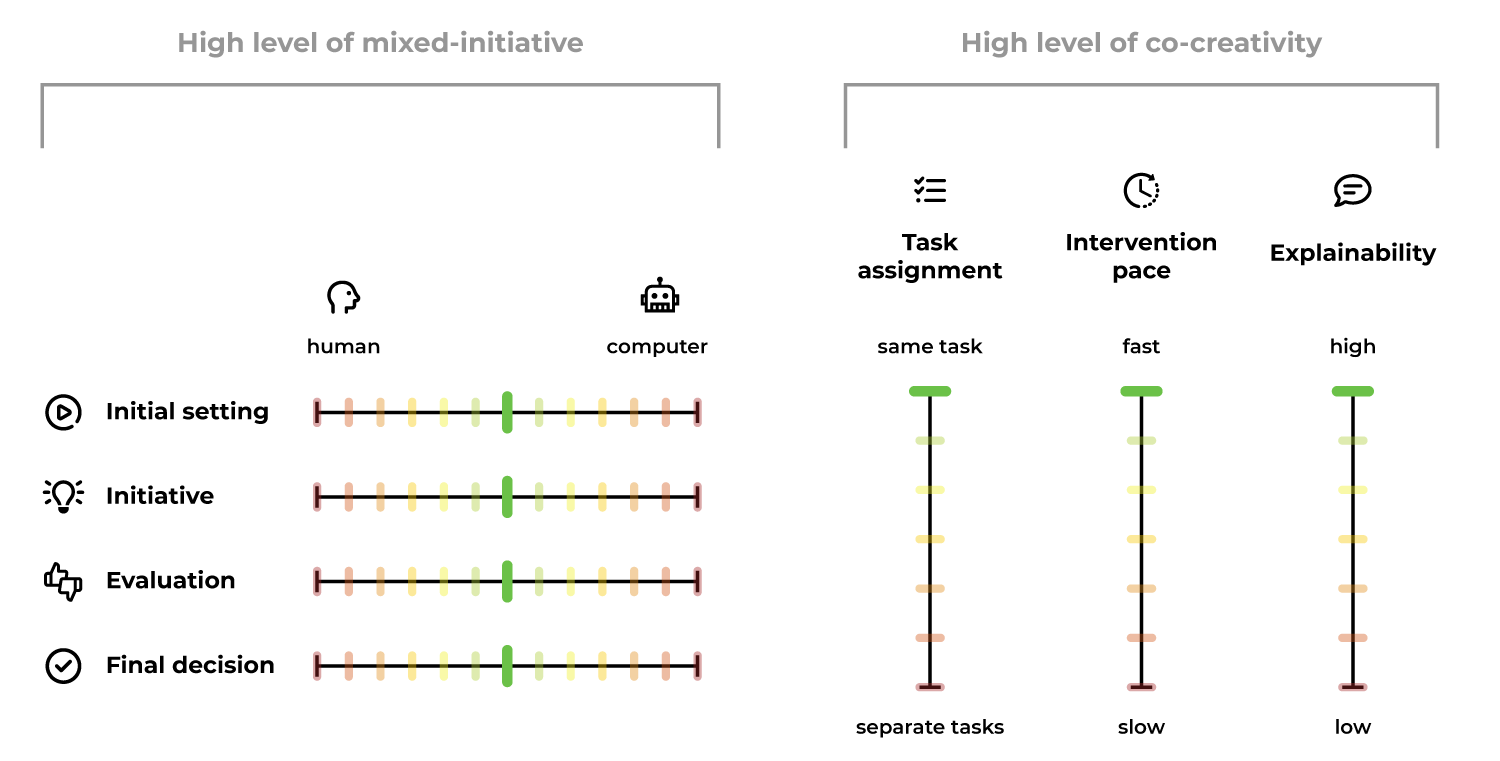}
\caption{Representation of a very high degree of mixed-initiative co-creativity through the use of MI-CCy Quantifier.}
\label{framework}
\end{figure}

With this perspective in mind, we present the MI-CCy Quantifier, a framework for comparing systems with respect to the degree of mixed-initiative co-creativity. The classification is made independently of the process flow. In this way, the analyzed parameters are generalized and applicable to any system, even if the processes are very different in their form. This model allows for a more direct visualization of which agent has a greater impact and how seamless their collaboration is. Figure~\ref{framework} shows the framework representing a scenario with a very high degree of mixed-initiative co-creativity. We emphasize that the MI-CCy Quantifier is not intended as a tool for rigorous classification, but rather as a means to support structured reflection on the mixed-initiative approach one seeks to develop.

We introduce four parameters that allow us to classify the influence that the human and the computer have on the creative process: \emph{Initial Setting}, \emph{Initiative}, \emph{Evaluation}, and \emph{Final Decision}. Each of these parameters is assessed along a spectrum, with the human at one end and the computer at the other. The spectrum is divided into 13 levels in total, covering 7 levels between the human end and the middle (perfect balance) and 7 levels between the middle and the AI end. We consider that the more equal (towards the middle of the spectrum), the higher the degree of mixed-initiative.
The framework has three more parameters that classify the approach on how much it promotes seamless collaboration (meaning that collaboration is facilitated, continuous, frictionless, and ideas are easily blended). They are \emph{Task Assignment}, \emph{Intervention Pace}, and \emph{Explainability}. Each of these parameters comprises its own scale, which is displayed vertically. The scale is divided into 7 levels. We consider that the higher up the scale, the more seamless the collaboration between the human and the machine will be and, consequently, the greater the degree of co-creativity.
These parameters are depicted in a visual scheme, where, in addition to the position on the scale of each criterion, we also establish a color chart for easier interpretation of the system's MI-CCy level. Thus, if the evaluated parameter is green, it is because it aligns with the ideal conditions for strong MI-CCy, while intermediate colors leading to red denote varying degrees of compliance with the criteria.

However, it should be noted that just because one system has a higher degree of mixed-initiative co-creativity over another does not mean that the approach is better than the other. It only tells us that the human agent and the computational agent are more equal partners and are equipped for a high level of collaboration. Such may or may not be desirable, depending on the intended applicability of the system. If the user already has an established idea and only wants help to execute it, they do not expect to have a system that wants to have a strong initiative in the conceptual phase of the process. Nevertheless, in this tutorial, we encourage researchers interested in developing new mixed-initiative co-creative systems to seek to implement an approach with a higher degree of MI-CCy since there is still little exploration in the current state of the art of the potential of the computer as a true colleague in the creative process. Only by providing the conditions for humans and artificial agents to collaborate in a seamless way can we properly assess how humans benefit from a more equitable partnership with AI systems. 

We anticipate that a close and balanced collaboration between humans and AI can increase mutual adaptability, foster creativity and user engagement, and improve the integration of user goals and ideas with system-generated content. However, we must also be mindful of the challenges it poses. To work effectively as colleagues, we need to provide clear communication channels and an intuitive interface, as well as manage initiative and control thoughtfully. Poorly implemented approaches can overwhelm the user or lead to a reduced sense of agency. For this reason, establishing an equal relationship must be a deliberate decision that is appropriate to the creative context. Moreover, humans and AI have different capabilities that must be equated in order to provide a peer status with reciprocal influence that still accommodates different competencies.

It is worth clarifying that throughout this tutorial, we refer to both human and computational agents as capable of learning and exhibiting creativity in a practical sense. Drawing on Boden’s \cite{Boden2004} taxonomy of creativity, we emphasize the importance of supporting transformational creativity in co-creative contexts, where either agent can contribute to altering the conceptual space or the creative direction. Our perspective aligns with the concept of symmetric alternating co-creativity \cite{Kantosalo2016}, in which both human and AI agents take turns in creating an artifact that satisfies both parties, and can use transformational creativity to handle conflicts and adapt to a new situation. While computational creativity and learning are fundamentally different from their human partners, we use these terms to describe behaviors that play comparable roles in MI-CCy — such as producing novel content, responding to feedback, or adapting to a partner — and do not imply cognitive equivalence between humans and machines. The MI-CCy Quantifier is thus meant to help reflect on how initiative is shared between agents and how easily they collaborate, regardless of the underlying nature of their reasoning.

\subsubsection{Initial Setting}
\begin{description}
\item[Definition] Setting the starting point for the creative process.

\item[Key points to consider] What will you work on? What problem do you want to solve? What is the initial concept or rule set? Think about who can make these decisions. The human? The computer? Both? Can they do it to the same extent?

\item[Illustration] Often, the initial setting is defined by the human creator when working with systems focused on supporting the user. In these cases, the user already has an established base concept and needs help to better develop it and turn it into an artifact \cite{Coutinho2024, Kreminski2020a}. In other cases, as in some approaches that use evolutionary algorithms, the creative process begins with an initial population generated by the system \cite{Gallotta2023, Liapis2012, Liapis2012a}. The user can only adapt the content from there, which, of course, substantially limits the creativity of the human.

\item[Additional remarks] Ideally, in a collaboration, the initial setting should have input from all participants in the creative process. We expect that, in a mixed-initiative approach, the value of this parameter will be somewhere along the spectrum, not so much at one end (human only) nor so much at the other (computer only). Imagine a group project in a game design course in which students are asked to develop a game by the end of the semester. In the first class, all the group members get together and try to contribute ideas to define the concept of the game. It may happen that the group discards most ideas, and just the idea of one participant is used. Nonetheless, all of them influenced the generation of the initial concept, even if only in the role of accepting or discarding ideas, in addition to stimulating the creativity of the group with their suggestions.
Even so, it is certain that one of the participants in the co-creative process will have to give the starting point to initiate the conversation. A system designed to be a colleague should leave open the possibility of who will make the first contribution(s). As an example, in Germinate \cite{Kreminski2020a}, the human is the one who defines the first rule set. We envision that the possibility of asking the computational agent to do the initial setting could be implemented with little effort.
\end{description}

\subsubsection{Initiative}

\begin{description}
\item[Definition] Having strong initiative throughout the creative process. 

\item[Key points to consider] We highlight four elements to consider when reflecting on the initiative of each co-creator: quantity of interventions, proactivity, input quality, and variety in the type of actions they can perform. Who is more involved? Can they contribute voluntarily and autonomously? Can their input be considered creative? In what ways can they intervene? We should reflect on these elements together.

\item[Illustration] We might have, for instance, a computational agent that continuously makes suggestions, but these can be considered too random, unoriginal, and not adaptable enough to the partner's feedback. Furthermore, the system may be limited to making suggestions without a chance to question high-level concepts or make decisions about the artifact being produced. In other cases, the interventions of an agent may be more about assessing what the colleague is doing than contributing creative input.

\item[Additional remarks] As the literature on human creativity and computational creativity shows, there are several ways for the human and the computer to actively participate in the process. This is evidenced by the frameworks of Spoto and Oleynik \cite{Spoto2017}, Muller et al. \cite{Muller2020} and Grabe et al. \cite{Grabe2022}, which consider several possible actions for both agents. With so many ways to manifest initiative, it can be difficult to pinpoint where the system falls along this spectrum. In human-computer co-creativity, it is ideal that both the human and the computer can actively and regularly intervene during the creative process and can do so in different ways that add value to the generation of the artifact(s). We must then contemplate whether both agents have the same opportunities for action.
\end{description} 

\subsubsection{Evaluation}

\begin{description}
\item[Definition] Assessing the suggestions and artifact(s) being created. 

\item[Key points to consider] Who judges if something is good or not? Who evaluates suggestions and has the power to accept or reject them? One can contribute ideas in the creative process, but do they have space to express their opinion on the ideas of others? If one does not like the output generated in the process, can they criticize it?

\item[Additional remarks] When we talk about evaluation, it is not limited to stating whether the artifact is viable and a possible solution to the problem (e.g., assessing if a level is playable and balanced). While this objective assessment is pivotal, it is insufficient in a creative process. It is also necessary to assess whether the artifact is creative, which may involve considerations of how novel, interesting, surprising, and valuable it is \cite{Boden2004,Canaan2018}. This is a more subjective assessment that can potentially lead to disagreements between the participants involved. What happens in the vast majority of current works is that, in the event of disagreements, the human agent is the one who has the power to decide on the course to be taken. In a mixed-initiative co-creative scenario, we expect both the human and the computer to evaluate the generated content, express their likes or dislikes, and try to reach a solution that satisfies both parties.
\end{description}

\subsubsection{Final Decision}
\begin{description}
\item[Definition] Deciding when to end the creative process. 

\item[Key points to consider] Typically, the process ends when one or both agents consider that the artifact is finished or the creative problem has been solved. Identify who can make this decision or who has more authority over it.

\item[Additional remarks] In the current state of the art, it is very unlikely that the computational agent has any say on this. The human agent is the only one reflecting if a final solution has been reached and deciding when to conclude the process. The computer can be included in this decision by expressing how it feels about the state of the creative process: ``I'm happy with this result. Are you?''. Alternatively, it might disagree with the intention of its co-creator to end the process: ``I don't feel like we have yet reached an ideal solution to our problem.''. Through such dialogue, the process will hopefully end when both partners agree that a final result has been achieved. 
\end{description} 

\subsubsection{Task Assignment}
\begin{description}
\item[Definition] Ranging from working on separate and distinct tasks to working on the same task. 

\item[Key points to consider] A mixed-initiative tool should be ranked further down the scale of Task Assignment if the human and the computer are working on different tasks, so that contributions resulting from one task only slightly affect the other. Higher up the scale must be systems where agents are working on the same task, and the input of one agent directly affects the subsequent intervention of its co-creator. It is also necessary to reflect on how the participants perform the tasks. They may be working on the same task, but do they perform the same type of actions? Can they make use of the same tools?

\item[Illustration] An illustrative case is the difference between interventions done by designing (where manipulation is more direct) or by setting rules and metrics. The human and the computer can both be working on a level design task in terms of layout. However, if one can arrange the level tiles while the other can only handle parameters, they are not working on the same task to the fullest extent.

\item[Additional remarks] This parameter directly correlates to the XAID (eXplainable AI for Designers) spectrum of domain overlap described by Zhu et al. \cite{Zhu2018}, which distinguishes off-task co-creativity from on-task co-creativity.  In on-task co-creative scenarios, agents apply the same tools to the same task. Rezwana and Maher \cite{Rezwana2021,Rezwana2022} include task distribution as an interaction component in their model, comprising same task and task divided. The Task Assignment criterion also relates to Kantosalo and Toivonen’s definition of alternating co-creativity in contrast to task-divided co-creativity \cite{Kantosalo2016}. They emphasize the partners’ concern to satisfy the requirements of both parties in alternating settings.  

Admittedly, it may or may not be advantageous to have participants in the creative process working on the same task. In human-human co-creativity, it is also common to have a clear task division, as one person may have more skills for certain aspects of game design and the partner for others. One may be better at drawing and therefore be in charge of game art, while another is concerned with building the characters and storyline of the game. However, if we intend to have a collaborative process in which participants foster the creativity of each other and seek to achieve a result that pleases both, then working on the same task as equals is ideal. It is usual to assume that the computer is incapable of being as creative as humans and to limit the type of tasks it can do because of that. We therefore hope that future approaches may counter this preconception.
\end{description}

\subsubsection{Intervention Pace}
\begin{description}
\item[Definition] Ranging from long design shifts to immediate feedback. 

\item[Key points to consider] Whether they are working on the same task or not, how long do co-creators wait to give and receive feedback? How much of a leap in content is there between interventions? A mixed-initiative system with a slow intervention pace is characterized by long design shifts and large content leaps. At the top of the scale are systems with a fast intervention pace, where feedback is immediate and there are no extreme leaps in the content created.

\item[Illustration] Short shifts do not necessarily mean short content leaps. A system may respond immediately to user feedback, but it can be too big of a jump between user-generated and system-generated content. In such a case, the response may skip several design steps ahead without allowing the colleague to intervene halfway through, which limits the discussion and gradual evolution of the artifact throughout the process.

 \item[Additional remarks]  Although this parameter should be considered independently of Task Assignment, working on the same task is more likely to allow for a faster intervention pace. The exchange of ideas is promoted if the stakeholders can respond almost instantly to additions and changes made to the creative artifact. The turns taken by the participants should be short and adapted to the interventions of each other so that both can express their opinions and preferences throughout the process. In fact, the agents involved may not even adhere to precisely defined shifts, being able to interrupt the partner who is designing if they deem it appropriate. This naturally raises considerations of practical judgment and sensibility towards one another, so that every co-creator has the opportunity to speak without anyone imposing too much on others with their interventions. In addition, there shouldn't be large gaps between content. If a designer is soon confronted with a finished artifact, they feel less room for exploring and discussing ideas, as the creative possibilities have already been narrowed.
 
 Rezwana and Maher’s \cite{Rezwana2021,Rezwana2022} interaction components of Participation Style and Timing of Initiative relate to this criterion. Participation Style distinguishes turn-taking from simultaneous contribution of both agents. Timing of Initiative distinguishes planned interventions from spontaneous ones. We consider there is a causal relationship between these two components, as simultaneous collaboration favors spontaneity while turn-taking limits it.

\end{description}

\subsubsection{Explainability}
\begin{description}
\item[Definition] Ranging from a low level of explainability to a high level of explainability. By explainability, we mean the ability of co-creators to explain their intentions, creative input, and process.

\item[Key points to consider] Are the agents involved in the process able to explain their ideas? Do they have any way of justifying their contributions? If so, in what way(s)? What strategies do they use? What are the means of communication? Is the creative partner able to understand? Can the partner also explain themself and be understood? Is it possible to solve through dialogue the conflicts that may arise? Do both agents have room to further defend ideas that have been rejected? Can they recall past interactions that are relevant to the present discussion? As we answer these questions, it becomes increasingly clear whether the mixed-initiative tool has sufficient explainability. A system with low explainability, and therefore at the bottom of the spectrum, does not have the means for co-creative agents to explain and understand the interventions of each other. A system with high explainability allows for a dialogue very close to what happens in human-human co-creativity. Incidentally, it may even be an improved version of this communication, thanks to audiovisual, memory, and processing capabilities that are unique to computational agents.

  \item[Additional remarks] Explanatory capabilities are critical for streamlining the collaborative process between humans and computers. As we know, communication is an intrinsic and essential element in human-human collaboration. Despite that, it is often overlooked in human-computer co-creativity. Several authors have identified the need to capacitate computational creativity systems with explainability, identifying requirements and strategies to do so. Zhu et al. \cite{Zhu2018} propose a subfield of Explainable Artificial Intelligence (XAI) focused on the specific needs of game designers when working with co-creative systems. Another subfield of XAI for contexts of computational creativity is proposed by Llano et al. \cite{Llano2020}. They introduce Explainable Computational Creativity (XCC), specifically concerned with enabling two-way communication between creative systems and human users. They identify four essential design principles for co-creative systems to achieve this: mental models, long-term memory, argumentation, and exposing the creative process. Margarido et al. \cite{Margarido2022} expand on how these principles can be applied to mixed-initiative co-creativity in game design, suggesting several approaches adapted to distinct use cases. Closely related to explainability, Lin et al. \cite{Lin2022} identify three non-exhaustive dimensions of communication between humans and computational agents: whether communication is human-initiated or agent-initiated, whether it concerns previous contributions (reflection) or future ones (elaboration), and whether it focuses on a general aspect of the creative task (global scope) or an individual one (local scope).
  
It is still challenging to achieve a high degree of explainability today. However, we incentivize researchers to make this attempt, as it will lead to a new paradigm for human-computer collaboration. We recommend relying on the literature to find directions on how to do so.
\end{description}

\subsection{Literature Analysis with MI-CCy Quantifier}\label{LiteratureClassification}

Using the MI-CCy Quantifier, we analyze below each of the works we selected in Section~\ref{MIGameDesign}, describing and justifying the classification we gave to each of the framework's parameters.

\subsubsection{Initial Setting}
\hfill

\textbf{AI Spaceship Generator (SG):} In this work, the initial setting is exclusively defined by the computational agent. The creative process begins with the system generating an initial population of spaceships, and the user forcibly works from any of these suggestions or requests the system to generate a new population. The conceptual space predefined by the system is too constrained to allow extensive exploration and cannot be transformed.

\textbf{Sprite Editor for Pixel Art Characters (SE):}  The human agent is the one who initiates the co-creative process, setting the starting point to generate other poses for the character. The computer cannot generate meaningful content without sufficiently comprehensive input from the human partner. Furthermore, if the initial design of the human agent is scarcely recognizable to the trained model, the quality of the computational agent's contributions is compromised.

\textbf{Sentient Sketchbook (SS):} The conceptual space is already predefined by the system. The tool is designed to generate maps for strategy games, meaning that the available tiles for constructing the map and the fitness dimensions used are already established for this purpose. Consequently, the human agent is constrained to work within a specific game style and towards a predetermined objective. As for initiating the co-creative process, the computational agent immediately starts offering suggestions, even if the user has not yet begun editing the map. On the other hand, the user can start drawing right away, triggering an instant change in the system's suggestions and thereby establishing the starting point for shaping the map. Taking all these factors into account, we consider that the initial setting criterion leans slightly towards the computer end of the spectrum due to the conceptual space being fully predefined.

\textbf{RL Brush (RB):} The conceptual space is predefined by the system. Since this version of the tool was specifically built to design levels for the Sokoban game, only the tiles necessary for this game are made available. The human agent is restricted to working with these tiles and cannot add new ones or change the conceptual space. The system also takes the first step in initiating the creative process. It invariably presents the user with an empty map containing a player tile at the center, along with four initial suggestions, which can slightly influence the direction in which the level will develop. However, since the human agent can immediately start editing the level, they can easily change this initial configuration, triggering new suggestions from the computational agent and taking control of the process.

\textbf{miWFC (mW):} In this case, the conceptual space is not restricted to any specific game genre, and the tool can prove useful for creating different kinds of 2D assets beyond maps. The computational agent offers a wide range of templates that the human agent can choose from, but the human agent also has the option to add their own custom templates. Despite the computer offering several alternatives for the input that will serve as the basis to generate the image, it is ultimately the human designer's decision on which one to use. The co-creative process can be initiated by the human agent by manually creating a desired area on the canvas and then requesting the computational agent to continue filling in the remaining space. Alternatively, the computational agent can start generating content from a blank canvas, but only upon request from the user. As such, although both agents can contribute to the initial concept, it is the human agent who retains more control and determines the specific problem they want to address.

\textbf{Pitako (P):} At the onset of the creative process, the system always presents a default background sprite and three default suggestions. However, the human agent can immediately add a new sprite, triggering further suggestions from the system. Consequently, while the human designer may be influenced and even choose to accept the initial suggestions, they retain control over the addition of the first sprite that truly begins to shape the game. It is worth noting that the sprites available for the human agent to add are limited to the behaviors and images found within the system's catalog. While the catalog is extensive, it still imposes creative limitations on the human partner. Considering all these factors, we posit that the initial setting criterion leans more toward the computer end of the spectrum.

\textbf{Story Designer (SD):} The co-creative process begins with the system consistently presenting the same base graph and immediately providing a set of suggestions. Consequently, the human agent is inevitably prompted to shape the narrative structure around this initial content. While it is possible for the human designer to erase the entire graph and start anew, the initial setting presented by the computational agent is most likely to influence their creative process. Regarding the conceptual space, although the nodes encapsulate a high degree of abstraction that allows for numerous creative possibilities to elaborate the narrative afterward, they are nonetheless rooted in tropes. Hence, they remain considerably constrained by the narrative conventions commonly prevalent in existing games.

\textbf{1001 Nights (1N):} The human agent initiates the creative process by being the first to write, setting the starting point from which the story will unfold. Nevertheless, the computational agent still exerts some influence on the initial setting. In addition to continuing the player's narrative, the AI model is prompted to focus on an ancient Middle Eastern story related to the keywords representing all valid weapons. This, coupled with the game's own narrative setting and the goal of generating these weapons, steers the story toward this particular literary genre.

\textbf{Germinate (G):} The tool is tailored for a style of gameplay with simple mechanics, yet it still allows for the expression of specific rhetorical intent. Specifically, the games created with Germinate can only be played with the mouse, lack a game map, and offer limited variety in the types of dynamics possible between entities and resources. Consequently, part of the conceptual space is constrained by the computational agent. However, the meaning assigned to the game, its goals, rules, and interactions are yet to be defined. Typically, it is the human agent who establishes the initial rule set. The system provides some example cards to get things started, but it is up to the user to determine which entities, resources, relationships, and triggers they want to add. The human agent can also request the computational agent to generate a random card, but ultimately, the initial concept of the game is mostly determined by the human agent.

\begin{figure}[h]
  \centering
  \includegraphics[width=\linewidth]{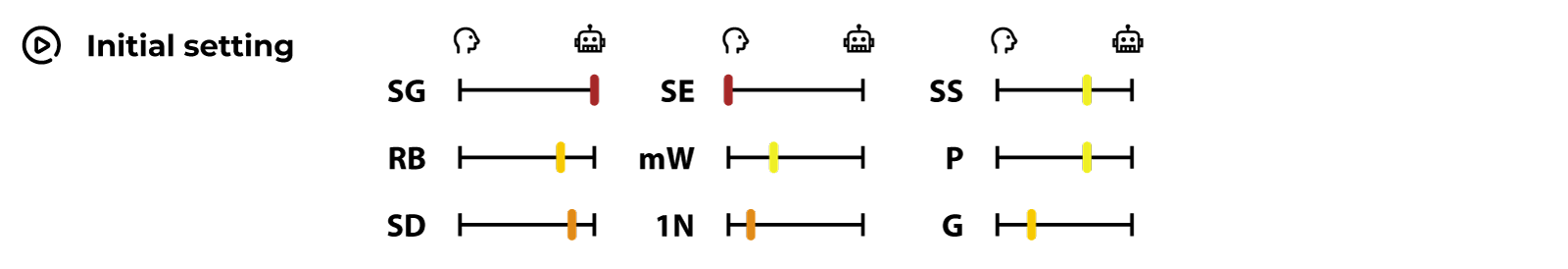}
\caption{Overview of the ranking of the Initial Setting for each work.}
\label{initial_setting}
\end{figure}

\subsubsection{Initiative}
\hfill

\textbf{AI Spaceship Generator (SG):} The computational agent can only intervene by generating several spaceship suggestions, while the user can only select one option and request new generation rounds. The user is the one who decides whether there will be a new round of spaceship generation and from what point (whether from the selected spaceship, a random spaceship, or resetting the population). However, the user has no way of directly modifying the artifacts or clearly indicating their preferred characteristics, restricting their creativity. The computational agent only generates more spaceships at the behest of its partner but can create their entire structure, exerting significantly greater control over the visual appearance of the artifacts. Thus, we consider the system to have more initiative.

\textbf{Sprite Editor for Pixel Art Characters (SE):} The human agent has the stronger initiative. They can directly edit any pose view, including those in the suggestion slots. By being able to draw and promptly modify any artifact, the human agent is granted more creative freedom. Meanwhile, the computer can only intervene upon user request, lacking proactivity to make independent contributions. It provides suggestions but cannot manipulate the artifacts directly. Moreover, the computational agent always tries to match the design provided by the human partner, avoiding the suggestion of images that are conceptually different from the source. Thus, its input lacks creativity.

\textbf{Sentient Sketchbook (SS):} The human agent contributes by directly editing the map sketch and selecting suggestions. The computational agent intervenes by consistently suggesting alternative map designs but cannot directly edit the main sketch. Both agents are highly proactive and engage multiple times throughout the creative process. The human partner can continuously add tiles and analyze suggestions, while the computer willingly responds with new creative input after each contribution made by its counterpart. Thus, both agents are deeply involved in the process and attentive to each other's feedback. Still, the user slightly stands out due to the ability to directly manipulate the map sketch.

\textbf{RL Brush (RB):} Similar to the Sentient Sketchbook, the human agent directly impacts the creative artifact throughout the process by being able to manually edit the level grid or choose a suggestion to replace it. The computational agent provides suggestions whenever the human partner makes a modification, but cannot directly edit the current level. This approach has the additional characteristic that the user can control the performance of the system, granting them the power to limit the quality and scope of its suggestions. Consequently, both agents proactively contribute creative input, but the human designer retains more control over the process.

\textbf{miWFC (mW):} The human agent holds more initiative throughout the creative process. The extent of contributions from the computational agent can vary significantly depending on how much the human designer allows it to intervene. Although the computational agent is practically capable of generating the entire creative artifact, it can only do so with permission from the human agent. Hence, the computational agent is not proactive. Furthermore, the system aims to generate an image similar to the initial input and aligned with the tile properties adjusted by the user. It does not actively seek to create content that is novel and surprising, thus being less creative than the human co-creator.

\textbf{Pitako (P):}  There are several actions that the human agent can perform. They can manually add sprites, accept sprite suggestions, manipulate various sprite properties, define interactions, accept interaction suggestions, position sprites on the game map, and define rules for game completion. On the other hand, the computational agent can only provide recommendations for sprites and interactions. Regarding sprite suggestions, the system autonomously provides them whenever the human designer adds a new sprite to the set. However, there comes a point when the sprite set becomes extensive, and the system can no longer generate new suggestions based on that set. As for interaction suggestions, these are presented only upon explicit user request. Thus, the human agent has considerably greater initiative. On top of that, the computer's contribution is limited in creativity as it relies on associations from existing games.

\textbf{Story Designer (SD):} During the co-creative process, the human agent retains the capability to directly edit the narrative graph, whereas the computational agent provides suggestions prompted by human interventions without explicit request. Although the computational agent contributes autonomously, with a focus on presenting novel and interesting content, it remains unable to manipulate the main graph. Furthermore, each suggestion only becomes visible if the user opts to inspect it. As a result, the system might offer multiple suggestions that the user may never come across. For these reasons, the human agent has the stronger initiative.

\textbf{1001 Nights (1N):} Both the human and the computational agent exhibit a strong initiative as they contribute equally by writing chunks of the narrative alternately. They are both attentive and adapt to each other's input in a creative manner. The only difference lies in the human co-creator having a limited number of characters for their writing turn, which allows the system to generate more extensive content.

\textbf{Germinate (G):} Once the human agent specifies their intention by defining the cards, the system generates a batch of playable games, each with its established rules and dynamics. If the human co-creator has allowed it beforehand, the system can also generate entities and resources for each game beyond those defined by the human partner. Throughout the co-creative process, the system recurrently generates more batches of games for the user to explore. After the human agent makes changes to the properties and requests the system to generate a new set of games, the current game list is cleared, and the system starts generating new games based on the updated settings. We consider the computational agent to have more initiative as it can continuously create various complete games, whereas the human agent can only define cards and cannot directly edit any game. Furthermore, despite the games generated by the system being constrained by the user's expressed intent, the system can typically add a fair amount of content to the game beyond what the user established. Therefore, the computational agent still retains a significant degree of creative freedom.

\begin{figure}[h]
  \centering
  \includegraphics[width=\linewidth]{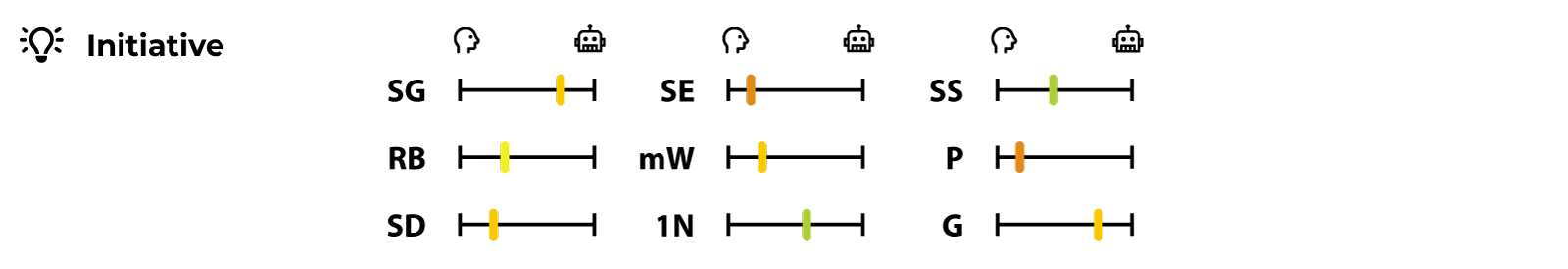}
\caption{Overview of the ranking of the Initiative for each work.}
\label{initiative}
\end{figure}

\subsubsection{Evaluation}
\hfill

\textbf{AI Spaceship Generator (SG):} The human agent evaluates the artifacts by selecting the one they prefer to use as a basis for a new iteration. However, the system can only predict the reasons that lead the human agent to prefer a particular spaceship over others. That is, the evaluation criteria of the human agent are assumed. In theory, the system favors user preferences and adjusts the fitness function accordingly, but this is not guaranteed in practice.

\textbf{Sprite Editor for Pixel Art Characters (SE):} The system provides suggestions that align with the input from the user without having an opinion about what is being drawn. The evaluation process is entirely carried out by the human agent, who determines whether to accept or ignore suggestions and modify the content if they deem it necessary.

\textbf{Sentient Sketchbook (SS):} Both evaluate the content being created. The human agent self-assesses the content they are creating, and the map suggestions of the system, possibly choosing to replace the current sketch or to draw inspiration from certain features of the suggestions. On the other hand, the computer tests the playability of the main map sketch and evaluates it based on gameplay properties related to game pace and player balance. It does the same to generate map designs and choose the best ones to present to the user. Additionally, half of the map designs generated by the system are evaluated in terms of novelty, and the system selects the most distinct maps to include in the suggestions. There is, however, a significant difference in the evaluation process carried out by the two agents: the system can only disclose its opinion to the human partner regarding gameplay characteristics. It cannot express whether it finds a particular map creative or not. Furthermore, regardless of what the computational agent informs the user about its evaluation, only the user can accept or reject suggestions. Therefore, the human agent has more evaluative power.

\textbf{RL Brush (RB):} The evaluation is mostly done by the human agent. They not only self-assess the content they create but also evaluate the suggestions provided by the computational agent, deciding whether any of them are good enough to be accepted. As for the computational agent, the internal evaluation it performs to generate suggestions remains unclear. We lack insight into the considerations that the AI models of the system have regarding the modifications they propose and how these suggestions may enhance the current level. Even if the system does conduct some form of evaluation, it cannot express it. Therefore, it is the human agent who guides the level design process, retaining complete control over the desired outcome for the creative artifact.

\textbf{miWFC (mW):} When the human agent is manually editing the canvas, the computational agent assesses the tile or set of tiles that the human partner intends to insert and may reject the edit if it deems it unfitting. However, this evaluation performed by the system focuses on the accuracy with respect to the example input provided for the generation, rather than the creativity of the human agent's contribution. In contrast, the human agent continuously evaluates the quality of the content created by both themself and the computational agent, revisiting previous stages and making changes if deemed necessary. Thus, the evaluation process is primarily conducted by the human agent. 

\textbf{Pitako (P):} The evaluation is primarily conducted by the human designer, who assesses their contributions and the game they are designing, as well as the system's recommendations. The computational agent solely self-evaluates its suggestions in terms of ``confidence'' and communicates this assessment to the human partner. A high confidence value indicates that the suggested element is common in existing games, while a low confidence value indicates that it is less common, making the suggestion more novel. However, the computational agent does not conduct that same evaluation for the sprites and interactions added by the user.

\textbf{Story Designer (SD):} Both parties evaluate the creative artifacts. The human agent self-assesses the content they create and the suggestions presented by the computational agent. Conversely, the computational agent evaluates both the main graph and the suggestions it generates in terms of feasibility, while also considering other metrics such as interestingness and coherence. Additionally, the computational agent discloses its assessment of the main narrative graph, the hovered suggestion, and the selected suggestion by displaying fitness, interestingness, and coherence scores for each. However, regardless of the computational agent's evaluation of the quality and creativity of the artifacts, only the human agent holds the authority to accept suggestions if they prefer them over the current graph.

\textbf{1001 Nights (1N):} Both agents assess what the other has written, choosing to adapt to their colleague's storyline and further develop their idea or change the direction or topic of the narrative. However, the human agent can assess the story as a whole. In contrast, the computational agent can only recall the last five inputs from both itself and the human co-creator. While they respond to each other's input thoughtfully, neither agent can critique the creativity and quality of the contributions from their partner.

\textbf{Germinate (G):} The evaluation is primarily conducted by the human agent. While the system can generate numerous games, it is the user who explores them to identify any with desirable characteristics. Based on this assessment, the user may modify their intent, leading to the discarding of all previous games to generate a new list. The only assessment made by the system is to check for contradictions in the cards defined by the user that would prevent game generation, without specifying the nature of these contradictions. Importantly, the computational agent does not provide any input or judgment on the meaning and creativity of the rhetorical intent of the human co-creator.

\begin{figure}[h]
  \centering
  \includegraphics[width=\linewidth]{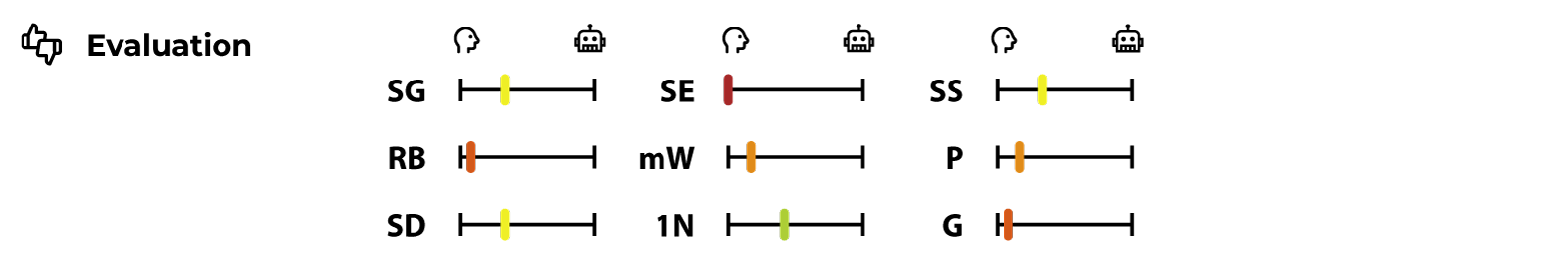}
\caption{Overview of the ranking of the Evaluation for each work.}
\label{evaluation}
\end{figure}

\subsubsection{Final decision}
\hfill

\textbf{AI Spaceship Generator (SG), Sprite Editor for Pixel Art Characters (SE), RL Brush (RB), miWFC (mW), Pitako (P), Story Designer (SD), Germinate (G):} The human agent is the one who decides when they are satisfied with an artifact (spaceship, character sprites, level map, texture, dynamics, narrative structure, or game, depending on the tool) and ready to end the creative process.

\textbf{Sentient Sketchbook (SS):} The human agent determines when they believe the map sketch is finalized and ready to be exported. Subsequently, the computational agent generates a detailed view of the map. The human partner cannot directly edit the detailed version of the map, but can choose their preferred option from the viewing modes provided by the computer or request the generation of a new view. Thus, the human has the final say over the structure of the level but has less control over its final look.

\textbf{1001 Nights (1N):} Since the narrative creation serves as a means to generate game mechanics, there is no aim for the story to reach a conclusion (unlike when storytelling itself is the primary goal). Nonetheless, it is the human agent who determines when they are ready to engage in battle with the King, thus concluding the co-creative storytelling session.

\begin{figure}[h]
  \centering
  \includegraphics[width=\linewidth]{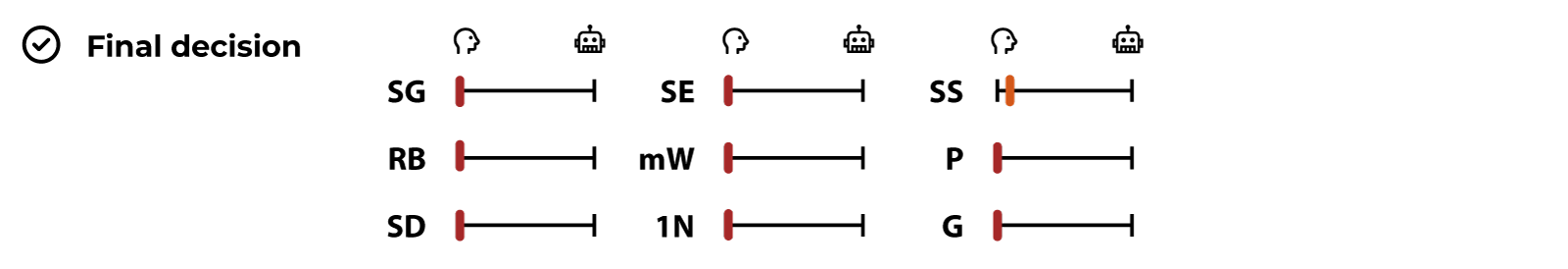}
\caption{Overview of the ranking of the Final decision for each work.}
\label{final_decision}
\end{figure}

\subsubsection{Task assignment}
\hfill

\textbf{AI Spaceship Generator (SG):} Even though both agents are working on creating spaceship designs, they do not work in the same way. The computational agent intervenes by generating the entire structure of the spaceships, while the human agent can only contribute by selecting a spaceship they prefer among those presented to them. This selection has no disruptive impact on the computer's subsequent intervention.

\textbf{Sprite Editor for Pixel Art Characters (SE):} Both agents are engaged in the creation of sprites for a pixel art character. The human agent has the ability to draw the character from scratch, directly manipulate any pose view, and select each image that will serve as input for a new suggestion. On the other hand, the computational agent can generate sprites for the character but can only present them as suggestions when prompted. In short, we can assert that both agents are able to generate content within the same domain, although not to the same degree.

\textbf{Sentient Sketchbook (SS):} The agents are collaborating on the shared task of creating map designs for a strategy game. The input from the human agent directly affects each subsequent intervention of the computational agent, whereas the contributions of the computational agent may or may not impact the next user intervention. Although the computer cannot edit the main sketch and select suggestions, it can generate map suggestions autonomously. Thus, both agents willingly construct map designs.

\textbf{RL Brush (RB):} Both agents are working on the shared task of designing a level for the Sokoban game. The contributions made by the human agent directly influence the suggestions presented by the computational agent. Conversely, the creative input from the computational agent may or may not affect the creative process carried out by the human co-creator. Lastly, even though the system cannot directly edit the grid of the current level, both agents contribute by adding or altering tiles.

\textbf{miWFC (mW):} The agents are involved in the same task of creating a 2D visual asset, such as a game level or texture. Both the human agent and the computational agent can directly edit the entire area of the bitmap, and their contributions strongly influence each other. However, there are some actions that the human agent can perform but the system cannot. Specifically, the human agent can delete content generated by the system and manipulate tile properties.

\textbf{Pitako (P):} The agents share some tasks in the creative process, but not all. Specifically, both are involved in creating sprites, mechanics, and dynamics. The human agent can directly add these elements to the game, whereas the computational agent only presents them as recommendations. Furthermore, only certain actions by the human creator directly impact the next contribution from the computer (e.g., adding a sprite with a behavior already present in the set does not trigger new suggestions). In the case of sprite placement on the level grid and the definition of termination rules, the computational agent does not contribute to these tasks.

\textbf{Story Designer (SD):} Both agents are engaged in the design of narrative structures and can produce the same type of content - narrative graphs. However, the human agent can manipulate the main graph by editing nodes and connections individually, whereas the computational agent contributes by presenting suggestions for complete graphs. The interventions made by the human agent directly influence the subsequent input of the computational agent, but the reverse may or may not occur.

\textbf{1001 Nights (1N):} Both the human and the computational agent are performing the same task and in the same manner. The creative work of one agent directly affects what their partner will write next.

\textbf{Germinate (G):} Both agents are focused on creating a game that expresses specific rhetorical goals. However, they cannot engage in the same type of actions in the co-creative process. While the computational agent is capable of generating complete games, the human agent can only define cards in a way that, hopefully, their intent is understood and conveyed in the games based solely on these guidelines. The contributions of the computational agent do not directly affect the subsequent actions of the human agent.

\begin{figure}[h]
  \centering
  \includegraphics[width=\linewidth]{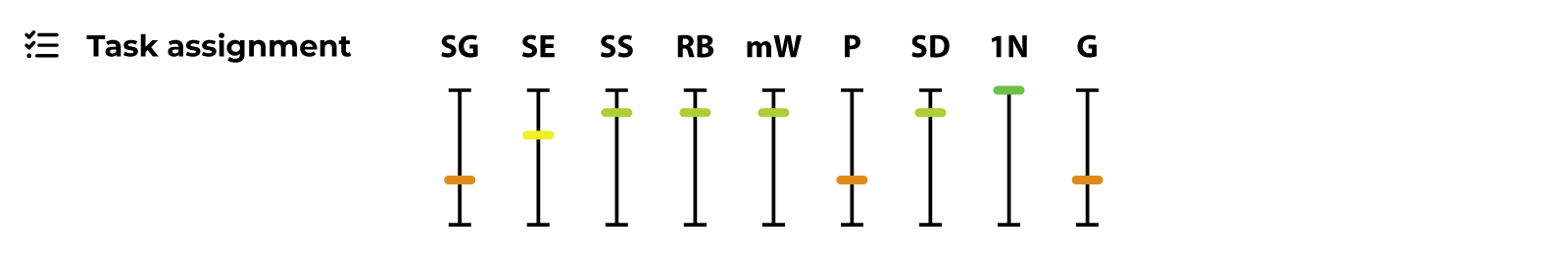}
\caption{Overview of the ranking of the Task assignment for each work.}
\label{task_assignment}
\end{figure}

\subsubsection{Intervention pace}
\hfill

\textbf{AI Spaceship Generator (SG):} The agents have clearly defined turns. Even if the user is quick to select one (or none) spaceship to evolve from in the next iteration, there is a long waiting period where the user cannot do anything nor receive feedback until the system completes the new generation. Furthermore, the system responds with new complete spaceship designs, as opposed to short modifications to the selected spaceship or more open-ended prototypes.

\textbf{Sprite Editor for Pixel Art Characters (SE):} The participants involved in the co-creative process can engage in a fast-paced interaction loop. Despite having well-defined turns for each participant (with the system only intervening upon user request), the response is immediate. In most cases, the suggestions from the system do not represent an extensive leap in content and are closely aligned with the user-generated content. Since the human agent has more control in the process, they can generate from none to a substantial amount of new content between the interventions of the computational agent. Although the turns are short, the system cannot intervene spontaneously.

\textbf{Sentient Sketchbook (SS):} The co-creators engage in a fast intervention pace, with short turns involving the addition of small portions of content over time. In this case, the human agent edits a single tile (or a small area) at a time, and the computational agent intervenes with each change made. The content leap between the main map sketch and the computer's suggestions is also not too substantial. There is only a larger gap observed at the beginning of the co-creative process as the system generates suggestions with sufficient content to be playable. In essence, the agents provide immediate and spontaneous feedback.

\textbf{RL Brush (RB):} The intervention pace of the computational agent is highly dependent on the values of the parameters ``step'' and ``tool radius''. When the ``step'' value is low and the ``tool radius'' is small, the feedback of the system is nearly instantaneous. However, if these values are increased, the system's response time also increases. Nevertheless, we consider this waiting time to be reasonable. Both the human and the computer edit one or a few tiles at a time, resulting in small content leaps. Consequently, the agents are not overwhelmed by a large amount of new information all at once, allowing them to gradually build the level and merge their ideas more easily.

\textbf{miWFC (mW):} The co-creators have well-defined shifts. When the human agent is editing the image, the computational agent waits and only intervenes upon the human partner's request. Despite this, the computational agent can generate content immediately, enabling rapid interaction between the agents. On the other hand, for the human designer to manually edit the canvas, they must switch screens, inevitably slowing down the intervention pace. The amount of content produced by each agent during their turn varies according to how the human agent chooses to interact with the computational agent. The computational agent can generate anything from a single tile to a complete bitmap, depending on the extent to which the human partner allows.

\textbf{Pitako (P):} The pace and spontaneity of co-creators' interventions depend on whether they are working on creating sprites or interactions. In the case of sprites, the turns are short, with the computational agent typically providing suggestions after each user intervention. Both parties focus on adding one sprite at a time. However, creating a sprite involves not only defining its graphical representation but also its behavior, which may require specifying several properties. Thus, we consider the content leap to be moderate. In the case of interactions, the system does not respond spontaneously to the contributions of the human creator, providing recommendations only upon request and solely based on the sprite set. Consequently, the user receives no feedback as they add interactions to the game.

\textbf{Story Designer (SD):} The system is not swift enough to generate suggestions in real-time as the user makes alterations, often resulting in delayed responses. Moreover, to view the suggestions generated by the computational agent, the human partner needs to inspect them one by one, thereby slowing down the intervention pace. As for the extent of the content leap, it varies from suggestion to suggestion. For instance, suggestions with a higher value in the ``step'' dimension will entail a more significant content leap.

\textbf{1001 Nights (1N):} In this approach, the agents have precise turns where each one contributes a portion of the story alternately. Although the turns are relatively short, we consider the content leap to be moderate, especially on the side of the computational agent, as it can generate more content in its turn. While the human agent may write one or two sentences, the computational agent can write an entire paragraph uninterrupted. The intervention pace would be faster if the agents could react while their partner is writing.

\textbf{Germinate (G):} Each agent has their own defined shift: the human agent is responsible for setting the cards, while the computational agent generates batches of games. These shifts tend to be lengthy. The system takes a considerable amount of time to generate the first batch of games, leaving the user waiting without any feedback during this period. Due to the extended duration of the computational agent's turn, it is likely that the human agent's turn will also be long, as they may choose to make multiple modifications at once. Moreover, as the content being created transitions from properties described in cards to full games generated by the system, the content leap is significant.

\begin{figure}[h]
  \centering
  \includegraphics[width=\linewidth]{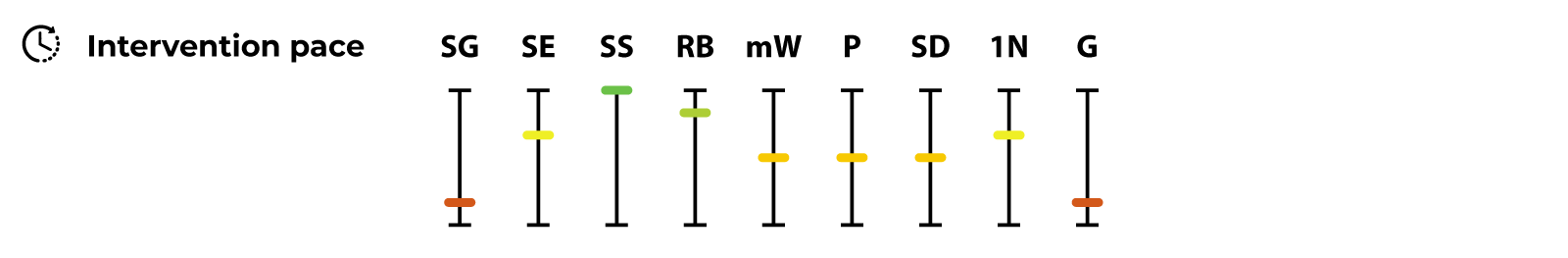}
\caption{Overview of the ranking of the Intervention pace for each work.}
\label{intervention_pace}
\end{figure}

\subsubsection{Explainability}
\hfill

\textbf{AI Spaceship Generator (SG), Sprite Editor for Pixel Art Characters (SE), RL Brush (RB):} There is no explainability. Neither the human nor the computer has the means to express their intentions and justify their contributions.

\textbf{Sentient Sketchbook (SS):} The explainability is severely limited. The computational agent makes a feeble attempt to justify why its suggestions may be preferable by displaying their respective fitness scores compared to the main map sketch. Thus, in the case of suggestions aimed at optimizing fitness scores, the human agent has a means to understand some of the improvements the suggestions offer to gameplay. However, these properties solely describe functional characteristics and provide no insight into the creativity of the map design. Conversely, the human designer lacks any means to explain their intentions and preferences.

\textbf{miWFC (mW):} There is little explainability between the human and the computational agent. All the user knows is that the system attempts to create content similar to the template used as an example, and according to the tile properties defined by the user. Additionally, the system's display of an animation of the generative process that the user can analyze step by step allows for a better understanding of the creative process of the computational agent. Nevertheless, the human agent has no other means of gaining further insight into the computational agent's decisions, nor does the computational agent have a way to comprehend the creative goals and preferences of the human agent.

\textbf{Pitako (P):} Explainability is almost non-existent. The computational agent provides a small hint regarding the novelty of its recommendations by displaying the confidence level, leaving it to the discretion of the human partner whether they want to create a game that closely resembles existing ones or opt for a set of sprites and interactions that is more distinct. However, it does not provide any further information about its creative process, and vice versa.

\textbf{Story Designer (SD):} Similar to the Sentient Sketchbook, the system displays the fitness, interestingness, and coherence scores of suggestions in relation to the main narrative graph. This provides some insight into the specific optimizations certain suggestions offer and why they might be preferable. However, this level of explainability remains insufficient, and there is no means for the human agent to reciprocally provide justification.

\textbf{1001 Nights (1N):} Since the game's objective is for the player to prompt the King to mention specific weapons, the player may sometimes become puzzled when the system's response lacks any of the keywords. In such cases, the system provides the player with a hint that they should tell a story more centered around swords and magic. However, this explanation is identical for every instance in which the King fails to mention any of the items.
It also provides no further insight into the overall creative process of the AI in story writing. Consequently, explainability is almost absent.

\textbf{Germinate (G):} This approach lacks extensive explainability, though it offers more than the previous works. In addition to using cards to express their intentions, the human co-creator can provide further clarification by adding tags to entity or resource cards. For instance, they may include a ``Stress'' resource with the tag ``player thinks it's bad'', enabling the computational agent to better grasp their vision. Conversely, each game generated by the system includes its rules displayed alongside in the form of cards. However, the presentation of these cards alone is often insufficient, and the human agent still struggles to comprehend the exact rules of the game and the message it aims to convey. A significant issue with this tool lies in its lack of explainability when it detects inconsistencies in the user's intent. In such cases, the system merely states that it cannot generate the games, without explaining the existing contradictions. This hinders the co-creative process, leaving the user unaware of what changes are necessary to resolve the issue.

\begin{figure}[h]
  \centering
  \includegraphics[width=\linewidth]{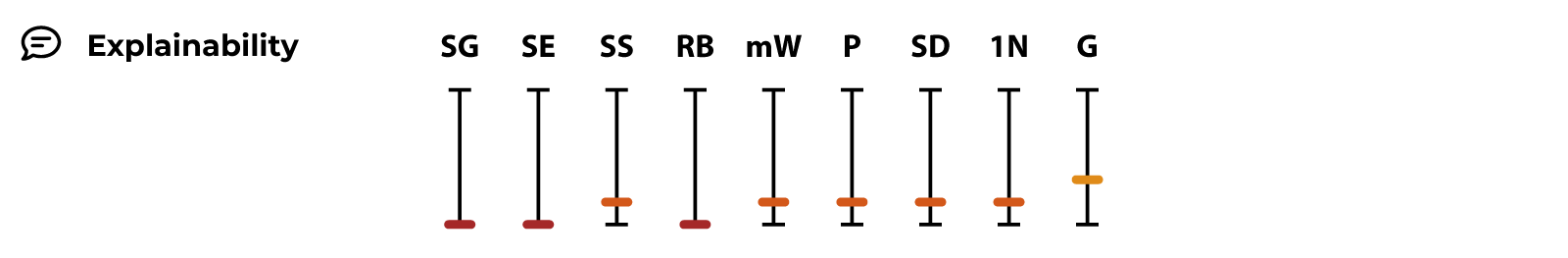}
\caption{Overview of the ranking of the Explainability for each work.}
\label{explainability}
\end{figure}

\subsubsection{Overview}

\begin{figure}[h]
  \centering
  \includegraphics[width=\linewidth]{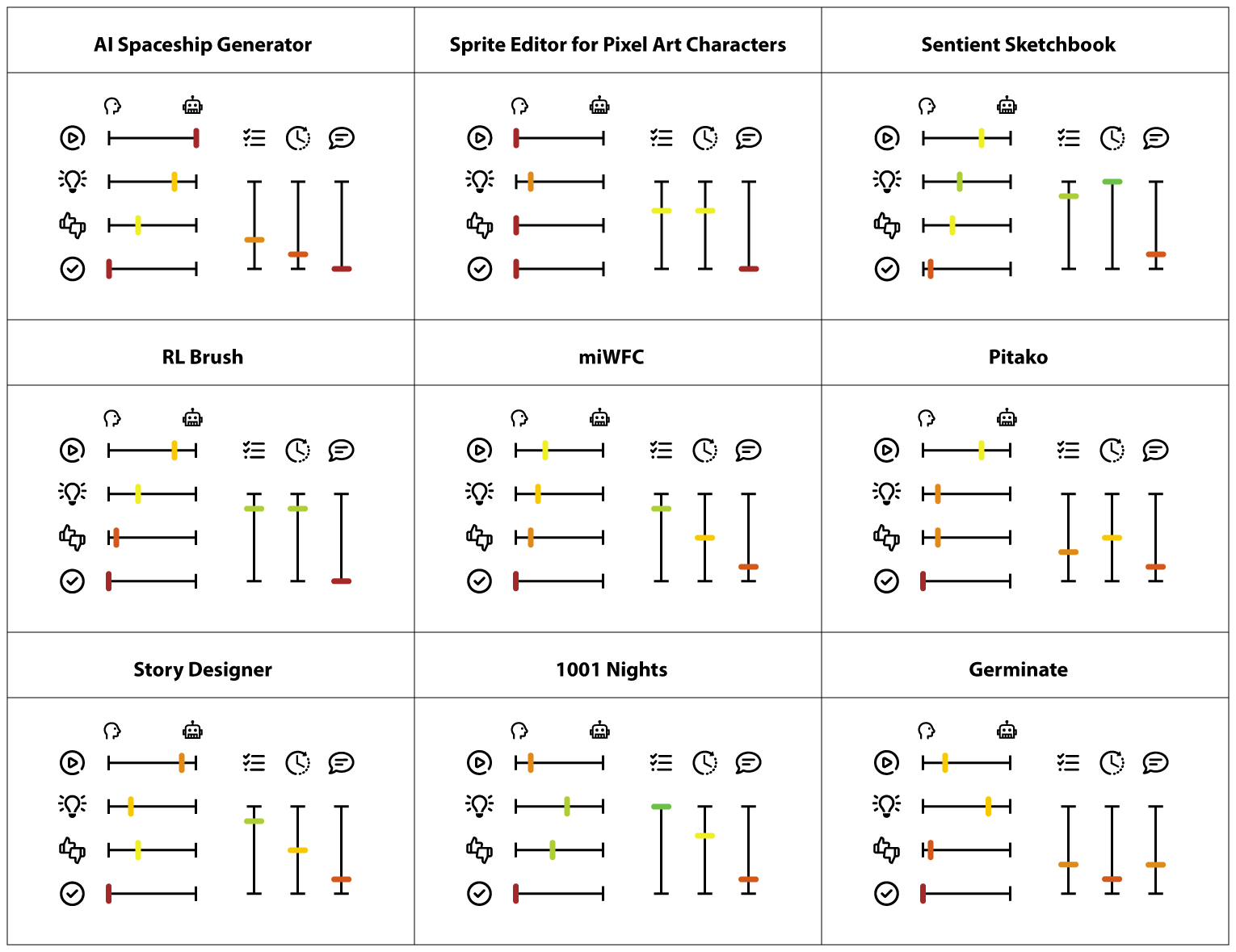}
\caption{Overview of the analysis of the degree of mixed-mixed-initiative co-creativity for each work using the MI-CCy Quantifier.}
\label{overview}
\end{figure}

As previously mentioned, this is not an exhaustive analysis, and different individuals may arrive at distinct conclusions regarding the level of MI-CCy in each tool. Nevertheless, it enables us to compare the approaches and identify relevant features in some that may be lacking in others. Additionally, it enables us to pinpoint characteristics common to most approaches, which can be addressed when developing future MI-CCy approaches in game design. In Figure~\ref{overview}, we provide an overview classification of each tool based on the MI-CCy Quantifier criteria. We can observe that the works that stand out with a higher level of MI-CCy are the Sentient Sketchbook and 1001 Nights. By examining Figures~\ref{initial_setting} to~\ref{explainability} or Figure~\ref{overview} alone, several patterns become discernible. Specifically, the initial setting tends to be skewed towards one of the agents. The evaluation of the process, artifacts, and creativity of the agents is primarily conducted by the human co-creator. Additionally, the final decision is always up to the human agent. Lastly, there is a general lack of explainability, both from the computational agent to the human agent and vice versa.

As mentioned earlier, game design is a broad discipline, so these approaches can address several types of game content with different complexities. Furthermore, there are a variety of techniques that can be used depending on the goals and intended functionality. We can identify trends in the techniques employed in these MI-CCy tools, such as the use of evolutionary algorithms. For instance, the Sentient Sketchbook uses genetic algorithms, while both the AI Spaceship Generator and Story Designer employ MAP-Elites. Machine learning algorithms are also prevalent, as seen in the RL Brush, which relies on reinforcement-learning-based models, and the Sprite Editor for Pixel Art Characters, which uses Generative Adversarial Networks. 1001 Nights makes use of Natural Language Processing, often seen in co-creative writing tools. Among the tools we have analyzed, there are also some less conventional content generation techniques. Namely, the miWFC applies the Wave Function Collapse algorithm, Germinate makes use of answer set programming to generate games, and Pitako applies a recommendation-based approach through frequent itemset mining.
The survey conducted by Lai et al. \cite{Lai2022} provides a more extensive review of the techniques used in mixed-initiative systems for game content creation. From our analysis, we contend that the underlying algorithms in these systems do not have as significant an impact on the level of MI-CCy as the means through which agents can contribute with creative input and communicate with one another. Nonetheless, we observe that techniques that encourage the computer to generate more novel and interesting content enable it to play a more significant role in the process.

In addition to the algorithms and techniques used, it is crucial to review how agents interact within the creative process. Regarding the computational agent, it is frequent for its contributions to involve presenting a set of suggestions from which the human agent can choose to accept one. This trend is observed in six out of the nine systems assessed using the MI-CCy Quantifier. Only in two works, namely miWFC and 1001 Nights, does the computational agent directly add content to the main creative artifact. Finally, in the Sprite Editor for Pixel Art Characters, the computational agent offers a single suggestion at a time for the user to accept or reject. Conversely, the human agent can add content directly to the final artifact in seven out of the nine frameworks, underscoring a clear distinction from their computational partner. The only exceptions are the AI Spaceship Generator, where the human agent is limited to selecting suggestions, and Germinate, where they can define cards instead of directly designing the game.

\section{Discussion and Open Issues}\label{Discussion}

The works presented in Section~\ref{MIGameDesign} and further explored in Section~\ref{LiteratureClassification} express a fair sample of the current progress in the development of mixed-initiative approaches to game design. Several other works were analyzed but not selected as use cases for this tutorial. Despite the diversity of recent works, there are some open issues. Below we describe the main ones identified:

\paragraph{Coarse collaboration}
Most approaches still fail to promote seamless collaboration between the human and the computer. In most cases, the human and the computational agent are still performing separate steps at a time and, therefore, not learning much from each other. For instance, in Germinate \citep{Kreminski2020a}, the user sets the initial properties, and then the computer generates games according to those defined properties. Subsequently, the user can analyze the generated content and make changes to the properties, followed by the computer generating more games. Various steps of modifying properties according to the new knowledge acquired from the output can result in a reasonable co-creative process. However, the rigid alternation of tasks remains noticeable, and the agents have a hard time engaging in a productive exchange of ideas that allows for the ideas to be smoothly blended.

\paragraph{Unbalanced partnership}
There is an unbalanced amount of initiative and creative input between the human and the computational agent. It is observable that some works favor computational initiative, while others favor human initiative. In the work of Gallotta et al. \cite{Gallotta2023}, the computational agent generates the first population of spaceships and has almost full control over what is presented. The human counterpart has low initiative and creative input, as they can only select spaceships throughout the process. In contrast, in the Sprite Editor for Pixel Art Characters \cite{Coutinho2024}, the computer is not as creative as the human, as the human partner creates the base character design, and the system always tries to match their vision during the course of co-creation. Compounding this issue, there are generally few instances where the computational agent is completely spontaneous, i.e., cases where it can react without requiring any trigger from the human counterpart. With the exception of when the computer presents an initial setting, if the human agent does not take any action,  the computer remains inactive as well.

\paragraph{Poor diversity of game content}
There is a lack of research in some domains of game design. Level design is the type of game content most tackled by current approaches. Although relevant, the high focus on map creation results in other critical aspects of game design receiving less consideration. Game content such as characters, sound, user interface, or complex interactions between game elements tends to be overlooked. Narrative design and the creation of visual assets, such as 2D or 3D models, although prevalent in many co-creative tools across broader domains, do not receive as much attention within the specific realm of video games.

\paragraph{Hardly generalizable}
The frameworks tend to be devised with a specific game genre and exploration space in mind, thus being difficult to generalize. In existing mixed-initiative tools, those focused on level design are typically restricted to a map style, such as dungeon crawler or platformer. Moreover, the visual aesthetics are often limited to what the framework provides, and the potential to expand it to other themes or art styles is unclear. Likewise, other content, such as narrative, mechanics, and interactions between game elements, are generally limited to a narrow solution space. An example is the level design approach in RL Brush \citep{Delarosa2021}, which is constrained to the particular design space of the Sokoban game. Consequently, these approaches are only suited for designing highly specific types of games.

\paragraph{No explainability}
Explainability is either nonexistent or severely limited. Few tools provide two-way communication channels between the agents involved in the creative process. In most cases, neither the human agent nor the computational agent has the means to explain and defend their ideas and contributions. In instances where the system does attempt to provide some form of explanation, it is typically confined to presenting some metric scores. If co-creative partners cannot comprehend each other's intentions and motivations, it becomes more challenging to converge their ideas and seamlessly integrate them into the final artifact.

\paragraph{Undefined evaluation methods}
Further research is required to establish appropriate methods for evaluating and validating these approaches. The evaluation of mixed-initiative approaches is a complex matter, given the multiple facets that must be considered, including the level of collaboration provided, the ability to foster creativity, and the quality and creative value of the content resulting from the design process. Nevertheless, this evaluation is necessary to rigorously compare different approaches and measure the progress made in the field.

\hfill

We recommend that individuals interested in developing novel mixed-initiative co-creative tools for game design should assess whether their system addresses any of these gaps. The MI-CCy Quantifier can serve as a valuable tool by providing a structured way to evaluate how well your developing system performs against each criterion and what improvements it brings to the current state of the art. Its utility is leveraged when the framework is used to analyze multiple co-creative systems. The comparison of how the systems differ across the MI-CCy Quantifier parameters allows us to identify trends and outliers, offering a measurable outlook on the challenges faced by current approaches.
Another point to highlight is that the analysis is always subjective to the individual experience of the user conducting it. If the MI-CCy Quantifier is used as a tool for individual reflection, this is not an issue. However, if the goal is to draw more objective conclusions about a system’s classification, a study should be conducted with multiple users to identify trends in their assessments.

We note that the MI-CCy Quantifier is not a closed framework, meaning that additional dimensions can potentially be incorporated beyond the ones we have discussed here. Usability is one such example. A user-friendly interface is crucial for facilitating interaction between the human and computational agents. While testing these tools, we observed variations in usability, with some demonstrating superior usability compared to others. This directly impacted our willingness to engage with the system and the quality of the co-creative process. A tool with poor usability generally results in a lower level of co-creativity.

On a final note, while this tutorial primarily focuses on providing guidelines for developing and assessing mixed-initiative co-creativity in the context of game design tools, we recognize that there are critical ethical issues surrounding AI usage that deserve attention. Some key concerns include how AI might affect intellectual property, especially as the division between human and computational-generated content becomes blurred, raising questions about authorship and ownership. There are also possible impacts on labor dynamics, as AI tools can change the roles and responsibilities of human creators, potentially affecting job security, fair compensation, and how human creativity is valued. On a broader level, AI systems might reinforce existing biases or reduce chances for human collaboration. These are complex issues that deserve thoughtful consideration, especially as AI becomes more involved in creative fields. Although this paper does not directly address these ethical challenges, we consider them important areas for future research to ensure that MI-CCy develops in a fair and responsible way.

\section{Conclusion}

In this paper, we provided guidance to researchers, developers, and game designers interested in implementing new approaches to game design focused on a high level of mixed-initiative co-creativity. To do so, we presented existing works on MI-CCy in game design targeted at several classes of game content. We introduced the MI-CCy Quantifier, a directive framework for analyzing works based on their level of MI-CCy in a descriptive and visually gradable manner. We illustrated its usage by applying it to nine of the presented frameworks. Through this analysis, we identified patterns in current approaches and characteristics that favor some over others. Furthermore, we highlight prevalent open issues in the state of the art, which we propose as key points to address when conceiving new MI-CCy frameworks for game design. In particular, we recommend that people working on forthcoming approaches focus on promoting a seamless and balanced partnership between the human agent and the computational agent. We also emphasize the need to address overlooked types of game content, to aim for greater versatility, and to enhance explainability. Finally, it is crucial to establish methods for evaluating MI-CCy approaches.

\begin{acks}
This work is partially financed through national funds by FCT -- Fundação para a Ciência e a Tecnologia, I.P., in the framework of the Project UIDB/00326/2025 and UIDP/00326/2025; and also supported by the Portuguese Recovery and Resilience Plan (PRR) through project C645008882-00000055, Center for Responsible AI.
The first author is funded by Fundação para a Ciência e a Tecnologia (FCT), Portugal, under the grant 2021.07047.BD.
\end{acks}

\bibliographystyle{ACM-Reference-Format}
\bibliography{collection}


\begin{thebibliography}{66}


\ifx \showCODEN    \undefined \def \showCODEN     #1{\unskip}     \fi
\ifx \showDOI      \undefined \def \showDOI       #1{#1}\fi
\ifx \showISBNx    \undefined \def \showISBNx     #1{\unskip}     \fi
\ifx \showISBNxiii \undefined \def \showISBNxiii  #1{\unskip}     \fi
\ifx \showISSN     \undefined \def \showISSN      #1{\unskip}     \fi
\ifx \showLCCN     \undefined \def \showLCCN      #1{\unskip}     \fi
\ifx \shownote     \undefined \def \shownote      #1{#1}          \fi
\ifx \showarticletitle \undefined \def \showarticletitle #1{#1}   \fi
\ifx \showURL      \undefined \def \showURL       {\relax}        \fi
\providecommand\bibfield[2]{#2}
\providecommand\bibinfo[2]{#2}
\providecommand\natexlab[1]{#1}
\providecommand\showeprint[2][]{arXiv:#2}

\bibitem[Alvarez et~al\mbox{.}(2018a)]%
        {Alvarez2018}
\bibfield{author}{\bibinfo{person}{Alberto Alvarez}, \bibinfo{person}{Steve Dahlskog}, \bibinfo{person}{Jose Font}, \bibinfo{person}{Johan Holmberg}, {and} \bibinfo{person}{Simon Johansson}.} \bibinfo{year}{2018}\natexlab{a}.
\newblock \showarticletitle{{Assessing aesthetic criteria in the evolutionary dungeon designer}}. In \bibinfo{booktitle}{\emph{Proceedings of the 13th International Conference on the Foundations of Digital Games}}. \bibinfo{publisher}{ACM}, \bibinfo{address}{New York, NY, USA}, \bibinfo{pages}{1--4}.
\newblock
\showISBNx{9781450365710}
\urldef\tempurl%
\url{https://doi.org/10.1145/3235765.3235810}
\showDOI{\tempurl}


\bibitem[Alvarez et~al\mbox{.}(2018b)]%
        {Alvarez2018a}
\bibfield{author}{\bibinfo{person}{Alberto Alvarez}, \bibinfo{person}{Steve Dahlskog}, \bibinfo{person}{Jose Font}, \bibinfo{person}{Johan Holmberg}, \bibinfo{person}{Chelsi Nolasco}, {and} \bibinfo{person}{Axel {\"{O}}sterman}.} \bibinfo{year}{2018}\natexlab{b}.
\newblock \showarticletitle{{Fostering creativity in the mixed-initiative evolutionary dungeon designer}}. In \bibinfo{booktitle}{\emph{Proceedings of the 13th International Conference on the Foundations of Digital Games}}. \bibinfo{publisher}{ACM}, \bibinfo{address}{New York, NY, USA}, \bibinfo{pages}{1--8}.
\newblock
\showISBNx{9781450365710}
\urldef\tempurl%
\url{https://doi.org/10.1145/3235765.3235815}
\showDOI{\tempurl}


\bibitem[Alvarez et~al\mbox{.}(2019)]%
        {Alvarez2019}
\bibfield{author}{\bibinfo{person}{Alberto Alvarez}, \bibinfo{person}{Steve Dahlskog}, \bibinfo{person}{Jose Font}, {and} \bibinfo{person}{Julian Togelius}.} \bibinfo{year}{2019}\natexlab{}.
\newblock \showarticletitle{{Empowering Quality Diversity in Dungeon Design with Interactive Constrained MAP-Elites}}. In \bibinfo{booktitle}{\emph{2019 IEEE Conference on Games (CoG)}}. \bibinfo{publisher}{IEEE}, \bibinfo{pages}{1--8}.
\newblock
\showISBNx{978-1-7281-1884-0}
\urldef\tempurl%
\url{https://doi.org/10.1109/CIG.2019.8848022}
\showDOI{\tempurl}


\bibitem[Alvarez et~al\mbox{.}(2022)]%
        {Alvarez2022}
\bibfield{author}{\bibinfo{person}{Alberto Alvarez}, \bibinfo{person}{Jose Font}, {and} \bibinfo{person}{Julian Togelius}.} \bibinfo{year}{2022}\natexlab{}.
\newblock \showarticletitle{{Story Designer: Towards a Mixed-Initiative Tool to Create Narrative Structures}}. In \bibinfo{booktitle}{\emph{Proceedings of the 17th International Conference on the Foundations of Digital Games}}. \bibinfo{publisher}{ACM}, \bibinfo{address}{New York, NY, USA}, \bibinfo{pages}{1--9}.
\newblock
\showISBNx{9781450397957}
\urldef\tempurl%
\url{https://doi.org/10.1145/3555858.3555929}
\showDOI{\tempurl}


\bibitem[Alvarez et~al\mbox{.}(2021)]%
        {Alvarez2021}
\bibfield{author}{\bibinfo{person}{Alberto Alvarez}, \bibinfo{person}{Eric Grevillius}, \bibinfo{person}{Elin Olsson}, {and} \bibinfo{person}{Jose Font}.} \bibinfo{year}{2021}\natexlab{}.
\newblock \showarticletitle{{Questgram [Qg]: Toward a Mixed-Initiative Quest Generation Tool}}. In \bibinfo{booktitle}{\emph{The 16th International Conference on the Foundations of Digital Games (FDG) 2021}}. \bibinfo{publisher}{ACM}, \bibinfo{address}{New York, NY, USA}, \bibinfo{pages}{1--10}.
\newblock
\showISBNx{9781450384223}
\urldef\tempurl%
\url{https://doi.org/10.1145/3472538.3472544}
\showDOI{\tempurl}


\bibitem[Baldwin et~al\mbox{.}(2017a)]%
        {Baldwin2017}
\bibfield{author}{\bibinfo{person}{Alexander Baldwin}, \bibinfo{person}{Steve Dahlskog}, \bibinfo{person}{Jose~M. Font}, {and} \bibinfo{person}{Johan Holmberg}.} \bibinfo{year}{2017}\natexlab{a}.
\newblock \showarticletitle{{Mixed-initiative procedural generation of dungeons using game design patterns}}. In \bibinfo{booktitle}{\emph{2017 IEEE Conference on Computational Intelligence and Games (CIG)}}. \bibinfo{publisher}{IEEE}, \bibinfo{pages}{25--32}.
\newblock
\showISBNx{978-1-5386-3233-8}
\urldef\tempurl%
\url{https://doi.org/10.1109/CIG.2017.8080411}
\showDOI{\tempurl}


\bibitem[Baldwin et~al\mbox{.}(2017b)]%
        {Baldwin2017a}
\bibfield{author}{\bibinfo{person}{Alexander Baldwin}, \bibinfo{person}{Steve Dahlskog}, \bibinfo{person}{Jose~M. Font}, {and} \bibinfo{person}{Johan Holmberg}.} \bibinfo{year}{2017}\natexlab{b}.
\newblock \showarticletitle{{Towards pattern-based mixed-initiative dungeon generation}}. In \bibinfo{booktitle}{\emph{Proceedings of the 12th International Conference on the Foundations of Digital Games}}. \bibinfo{publisher}{ACM}, \bibinfo{address}{New York, NY, USA}, \bibinfo{pages}{1--10}.
\newblock
\showISBNx{9781450353199}
\urldef\tempurl%
\url{https://doi.org/10.1145/3102071.3110572}
\showDOI{\tempurl}


\bibitem[Bhaumik et~al\mbox{.}(2021)]%
        {Bhaumik2021}
\bibfield{author}{\bibinfo{person}{Debosmita Bhaumik}, \bibinfo{person}{Ahmed Khalifa}, {and} \bibinfo{person}{Julian Togelius}.} \bibinfo{year}{2021}\natexlab{}.
\newblock \showarticletitle{{Lode Encoder: AI-constrained co-creativity}}. In \bibinfo{booktitle}{\emph{2021 IEEE Conference on Games (CoG)}}. \bibinfo{publisher}{IEEE}, \bibinfo{pages}{01--08}.
\newblock
\showISBNx{978-1-6654-3886-5}
\urldef\tempurl%
\url{https://doi.org/10.1109/CoG52621.2021.9619009}
\showDOI{\tempurl}


\bibitem[Boden(2004)]%
        {Boden2004}
\bibfield{author}{\bibinfo{person}{Margaret~A. Boden}.} \bibinfo{year}{2004}\natexlab{}.
\newblock \bibinfo{booktitle}{\emph{{The Creative Mind}}}.
\newblock \bibinfo{publisher}{Routledge}. 1--344 pages.
\newblock
\showISBNx{9781134379583}
\urldef\tempurl%
\url{https://doi.org/10.4324/9780203508527}
\showDOI{\tempurl}


\bibitem[Brown et~al\mbox{.}(2020)]%
        {Brown2020}
\bibfield{author}{\bibinfo{person}{Tom~B. Brown}, \bibinfo{person}{Benjamin Mann}, \bibinfo{person}{Nick Ryder}, \bibinfo{person}{Melanie Subbiah}, \bibinfo{person}{Jared Kaplan}, \bibinfo{person}{Prafulla Dhariwal}, \bibinfo{person}{Arvind Neelakantan}, \bibinfo{person}{Pranav Shyam}, \bibinfo{person}{Girish Sastry}, \bibinfo{person}{Amanda Askell}, \bibinfo{person}{Sandhini Agarwal}, \bibinfo{person}{Ariel Herbert-Voss}, \bibinfo{person}{Gretchen Krueger}, \bibinfo{person}{Tom Henighan}, \bibinfo{person}{Rewon Child}, \bibinfo{person}{Aditya Ramesh}, \bibinfo{person}{Daniel~M. Ziegler}, \bibinfo{person}{Jeffrey Wu}, \bibinfo{person}{Clemens Winter}, \bibinfo{person}{Christopher Hesse}, \bibinfo{person}{Mark Chen}, \bibinfo{person}{Eric Sigler}, \bibinfo{person}{Mateusz Litwin}, \bibinfo{person}{Scott Gray}, \bibinfo{person}{Benjamin Chess}, \bibinfo{person}{Jack Clark}, \bibinfo{person}{Christopher Berner}, \bibinfo{person}{Sam McCandlish}, \bibinfo{person}{Alec Radford}, \bibinfo{person}{Ilya Sutskever},
  {and} \bibinfo{person}{Dario Amodei}.} \bibinfo{year}{2020}\natexlab{}.
\newblock \showarticletitle{{Language models are few-shot learners}}.
\newblock \bibinfo{journal}{\emph{Advances in Neural Information Processing Systems}}  \bibinfo{volume}{33} (\bibinfo{year}{2020}).
\newblock
\showISSN{10495258}
\showeprint[arxiv]{2005.14165}


\bibitem[Butler et~al\mbox{.}(2013)]%
        {Butler2013}
\bibfield{author}{\bibinfo{person}{Eric Butler}, \bibinfo{person}{Adam~M. Smith}, \bibinfo{person}{Yun-En Liu}, {and} \bibinfo{person}{Zoran Popovic}.} \bibinfo{year}{2013}\natexlab{}.
\newblock \showarticletitle{{A mixed-initiative tool for designing level progressions in games}}. In \bibinfo{booktitle}{\emph{Proceedings of the 26th annual ACM symposium on User interface software and technology}}. \bibinfo{publisher}{ACM}, \bibinfo{address}{New York, NY, USA}, \bibinfo{pages}{377--386}.
\newblock
\showISBNx{9781450322683}
\urldef\tempurl%
\url{https://doi.org/10.1145/2501988.2502011}
\showDOI{\tempurl}


\bibitem[Canaan et~al\mbox{.}(2018)]%
        {Canaan2018}
\bibfield{author}{\bibinfo{person}{Rodrigo Canaan}, \bibinfo{person}{Stefan Menzel}, \bibinfo{person}{Julian Togelius}, {and} \bibinfo{person}{Andy Nealen}.} \bibinfo{year}{2018}\natexlab{}.
\newblock \showarticletitle{{Towards Game-based Metrics for Computational Co-Creativity}}. In \bibinfo{booktitle}{\emph{2018 IEEE Conference on Computational Intelligence and Games (CIG)}}. \bibinfo{publisher}{IEEE}, \bibinfo{pages}{1--8}.
\newblock
\showISBNx{978-1-5386-4359-4}
\urldef\tempurl%
\url{https://doi.org/10.1109/CIG.2018.8490429}
\showDOI{\tempurl}


\bibitem[Carbornell(1970)]%
        {Carbornell1970}
\bibfield{author}{\bibinfo{person}{Jaime~R. Carbornell}.} \bibinfo{year}{1970}\natexlab{}.
\newblock \showarticletitle{{Mixed-initiative man-computer instructional dialogues.}}
\newblock  (\bibinfo{year}{1970}).
\newblock


\bibitem[Chakrabarty et~al\mbox{.}(2022)]%
        {Chakrabarty2022}
\bibfield{author}{\bibinfo{person}{Tuhin Chakrabarty}, \bibinfo{person}{Vishakh Padmakumar}, {and} \bibinfo{person}{He He}.} \bibinfo{year}{2022}\natexlab{}.
\newblock \showarticletitle{{Help me write a Poem: Instruction Tuning as a Vehicle for Collaborative Poetry Writing}}. In \bibinfo{booktitle}{\emph{Proceedings of the 2022 Conference on Empirical Methods in Natural Language Processing}}. \bibinfo{publisher}{Association for Computational Linguistics}, \bibinfo{address}{Stroudsburg, PA, USA}, \bibinfo{pages}{6848--6863}.
\newblock
\urldef\tempurl%
\url{https://doi.org/10.18653/v1/2022.emnlp-main.460}
\showDOI{\tempurl}


\bibitem[Charity et~al\mbox{.}(2022)]%
        {Charity2022}
\bibfield{author}{\bibinfo{person}{M Charity}, \bibinfo{person}{Isha Dave}, \bibinfo{person}{Ahmed Khalifa}, {and} \bibinfo{person}{Julian Togelius}.} \bibinfo{year}{2022}\natexlab{}.
\newblock \showarticletitle{{Baba is Y'all 2.0: Design and Investigation of a Collaborative Mixed-Initiative System}}.
\newblock \bibinfo{journal}{\emph{IEEE Transactions on Games}} (\bibinfo{year}{2022}), \bibinfo{pages}{1--15}.
\newblock
\showISSN{2475-1502}
\urldef\tempurl%
\url{https://doi.org/10.1109/TG.2022.3223527}
\showDOI{\tempurl}


\bibitem[Charity et~al\mbox{.}(2020)]%
        {Charity2020}
\bibfield{author}{\bibinfo{person}{Megan Charity}, \bibinfo{person}{Ahmed Khalifa}, {and} \bibinfo{person}{Julian Togelius}.} \bibinfo{year}{2020}\natexlab{}.
\newblock \showarticletitle{{Baba is Y'all: Collaborative Mixed-Initiative Level Design}}. In \bibinfo{booktitle}{\emph{2020 IEEE Conference on Games (CoG)}}. \bibinfo{publisher}{IEEE}, \bibinfo{pages}{542--549}.
\newblock
\showISBNx{978-1-7281-4533-4}
\urldef\tempurl%
\url{https://doi.org/10.1109/CoG47356.2020.9231807}
\showDOI{\tempurl}


\bibitem[Coutinho and Chaimowicz(2024)]%
        {Coutinho2024}
\bibfield{author}{\bibinfo{person}{Fl{\'{a}}vio Coutinho} {and} \bibinfo{person}{Luiz Chaimowicz}.} \bibinfo{year}{2024}\natexlab{}.
\newblock \showarticletitle{{Pixel art character generation as an image-to-image translation problem using GANs}}.
\newblock \bibinfo{journal}{\emph{Graphical Models}}  \bibinfo{volume}{132} (\bibinfo{date}{apr} \bibinfo{year}{2024}), \bibinfo{pages}{101213}.
\newblock
\showISSN{15240703}
\urldef\tempurl%
\url{https://doi.org/10.1016/j.gmod.2024.101213}
\showDOI{\tempurl}


\bibitem[Dahlskog et~al\mbox{.}(2015)]%
        {Dahlskog2015}
\bibfield{author}{\bibinfo{person}{Steve Dahlskog}, \bibinfo{person}{Staffan Bj{\"{o}}rk}, {and} \bibinfo{person}{Julian Togelius}.} \bibinfo{year}{2015}\natexlab{}.
\newblock \showarticletitle{{Patterns, Dungeons and Generators}}. In \bibinfo{booktitle}{\emph{10th International Conference on the Foundations of Digital Games}}.
\newblock


\bibitem[Davis(2013)]%
        {Davis2013}
\bibfield{author}{\bibinfo{person}{Nicholas Davis}.} \bibinfo{year}{2013}\natexlab{}.
\newblock \showarticletitle{{Human-computer co-creativity: blending human and computational creativity}}. In \bibinfo{booktitle}{\emph{Proceedings of the Ninth Artificial Intelligence and Interactive Digital Entertainment Conference}}.
\newblock


\bibitem[Delarosa et~al\mbox{.}(2021)]%
        {Delarosa2021}
\bibfield{author}{\bibinfo{person}{Omar Delarosa}, \bibinfo{person}{Hang Dong}, \bibinfo{person}{Mindy Ruan}, \bibinfo{person}{Ahmed Khalifa}, {and} \bibinfo{person}{Julian Togelius}.} \bibinfo{year}{2021}\natexlab{}.
\newblock \showarticletitle{{Mixed-Initiative Level Design with RL Brush}}. In \bibinfo{booktitle}{\emph{Artificial Intelligence in Music, Sound, Art and Design. EvoMUSART 2021}}. \bibinfo{publisher}{Springer International Publishing}, \bibinfo{pages}{412--426}.
\newblock
\urldef\tempurl%
\url{https://doi.org/10.1007/978-3-030-72914-1_27}
\showDOI{\tempurl}


\bibitem[Deterding et~al\mbox{.}(2017)]%
        {Deterding2017}
\bibfield{author}{\bibinfo{person}{Sebastian Deterding}, \bibinfo{person}{Jonathan Hook}, \bibinfo{person}{Rebecca Fiebrink}, \bibinfo{person}{Marco Gillies}, \bibinfo{person}{Jeremy Gow}, \bibinfo{person}{Memo Akten}, \bibinfo{person}{Gillian Smith}, \bibinfo{person}{Antonios Liapis}, {and} \bibinfo{person}{Kate Compton}.} \bibinfo{year}{2017}\natexlab{}.
\newblock \showarticletitle{{Mixed-Initiative Creative Interfaces}}. In \bibinfo{booktitle}{\emph{Proceedings of the 2017 CHI Conference Extended Abstracts on Human Factors in Computing Systems}}. \bibinfo{publisher}{ACM}, \bibinfo{address}{New York, NY, USA}, \bibinfo{pages}{628--635}.
\newblock
\showISBNx{9781450346566}
\urldef\tempurl%
\url{https://doi.org/10.1145/3027063.3027072}
\showDOI{\tempurl}


\bibitem[Dias et~al\mbox{.}(2024)]%
        {Dias2024}
\bibfield{author}{\bibinfo{person}{Mariana Dias}, \bibinfo{person}{Pedro Coelho}, \bibinfo{person}{Rui Figueiredo}, \bibinfo{person}{Rita Carvalho}, \bibinfo{person}{Veronica Orvalho}, {and} \bibinfo{person}{Alexis Roche}.} \bibinfo{year}{2024}\natexlab{}.
\newblock \showarticletitle{{Creating Infinite Characters From a Single Template: How Automation May Give Super Powers to 3D Artists}}. In \bibinfo{booktitle}{\emph{ACM SIGGRAPH 2024 Talks}}. \bibinfo{publisher}{ACM}, \bibinfo{address}{New York, NY, USA}, \bibinfo{pages}{1--2}.
\newblock
\showISBNx{9798400705151}
\urldef\tempurl%
\url{https://doi.org/10.1145/3641233.3664339}
\showDOI{\tempurl}


\bibitem[Gallotta et~al\mbox{.}(2023)]%
        {Gallotta2023}
\bibfield{author}{\bibinfo{person}{Roberto Gallotta}, \bibinfo{person}{Kai Arulkumaran}, {and} \bibinfo{person}{L.~B. Soros}.} \bibinfo{year}{2023}\natexlab{}.
\newblock \showarticletitle{{Preference-Learning Emitters for Mixed-Initiative Quality-Diversity Algorithms}}.
\newblock \bibinfo{journal}{\emph{IEEE Transactions on Games}} (\bibinfo{year}{2023}), \bibinfo{pages}{1--14}.
\newblock
\showISSN{2475-1502}
\urldef\tempurl%
\url{https://doi.org/10.1109/TG.2023.3264457}
\showDOI{\tempurl}


\bibitem[{Gon{\c{c}}alo Oliveira} et~al\mbox{.}(2017)]%
        {Oliveira2017}
\bibfield{author}{\bibinfo{person}{Hugo {Gon{\c{c}}alo Oliveira}}, \bibinfo{person}{Tiago Mendes}, {and} \bibinfo{person}{Ana Boavida}.} \bibinfo{year}{2017}\natexlab{}.
\newblock \showarticletitle{{Co-PoeTryMe: a Co-Creative Interface for the Composition of Poetry}}. In \bibinfo{booktitle}{\emph{Proceedings of the 10th International Conference on Natural Language Generation}}. \bibinfo{publisher}{Association for Computational Linguistics}, \bibinfo{address}{Stroudsburg, PA, USA}, \bibinfo{pages}{70--71}.
\newblock
\urldef\tempurl%
\url{https://doi.org/10.18653/v1/W17-3508}
\showDOI{\tempurl}


\bibitem[Grabe et~al\mbox{.}(2022)]%
        {Grabe2022}
\bibfield{author}{\bibinfo{person}{Imke Grabe}, \bibinfo{person}{Miguel Gonz{\'{a}}lez-Duque}, \bibinfo{person}{Sebastian Risi}, {and} \bibinfo{person}{Jichen Zhu}.} \bibinfo{year}{2022}\natexlab{}.
\newblock \showarticletitle{{Towards a Framework for Human-AI Interaction Patterns in Co-Creative GAN Applications}}. In \bibinfo{booktitle}{\emph{Proceeding of the 3rd Workshop on Human-AI Co-Creation with Generative Models (HAI-GEN ‘22) at ACM IUI Workshops}}, Vol.~\bibinfo{volume}{3124}.
\newblock
\showISSN{16130073}


\bibitem[Guzdial et~al\mbox{.}(2019)]%
        {Guzdial2019}
\bibfield{author}{\bibinfo{person}{Matthew Guzdial}, \bibinfo{person}{Nicholas Liao}, \bibinfo{person}{Jonathan Chen}, \bibinfo{person}{Shao-Yu Chen}, \bibinfo{person}{Shukan Shah}, \bibinfo{person}{Vishwa Shah}, \bibinfo{person}{Joshua Reno}, \bibinfo{person}{Gillian Smith}, {and} \bibinfo{person}{Mark~O. Riedl}.} \bibinfo{year}{2019}\natexlab{}.
\newblock \showarticletitle{{Friend, Collaborator, Student, Manager}}. In \bibinfo{booktitle}{\emph{Proceedings of the 2019 CHI Conference on Human Factors in Computing Systems}}. \bibinfo{publisher}{ACM}, \bibinfo{address}{New York, NY, USA}, \bibinfo{pages}{1--13}.
\newblock
\showISBNx{9781450359702}
\urldef\tempurl%
\url{https://doi.org/10.1145/3290605.3300854}
\showDOI{\tempurl}


\bibitem[Hendrikx et~al\mbox{.}(2013)]%
        {Hendrikx2013}
\bibfield{author}{\bibinfo{person}{Mark Hendrikx}, \bibinfo{person}{Sebastiaan Meijer}, \bibinfo{person}{Joeri {Van Der Velden}}, {and} \bibinfo{person}{Alexandru Iosup}.} \bibinfo{year}{2013}\natexlab{}.
\newblock \showarticletitle{{Procedural content generation for games}}.
\newblock \bibinfo{journal}{\emph{ACM Transactions on Multimedia Computing, Communications, and Applications}} \bibinfo{volume}{9}, \bibinfo{number}{1} (\bibinfo{date}{feb} \bibinfo{year}{2013}), \bibinfo{pages}{1--22}.
\newblock
\showISSN{1551-6857}
\urldef\tempurl%
\url{https://doi.org/10.1145/2422956.2422957}
\showDOI{\tempurl}


\bibitem[Kantosalo and Toivonen(2016)]%
        {Kantosalo2016}
\bibfield{author}{\bibinfo{person}{Anna Kantosalo} {and} \bibinfo{person}{Hannu Toivonen}.} \bibinfo{year}{2016}\natexlab{}.
\newblock \showarticletitle{{Modes for creative human-computer collaboration: alternating and task-divided co-creativity}}. In \bibinfo{booktitle}{\emph{Proceedings of the Seventh International Conference on Computational Creativity}}. \bibinfo{address}{Paris}.
\newblock


\bibitem[Karavolos et~al\mbox{.}(2015)]%
        {Karavolos2015}
\bibfield{author}{\bibinfo{person}{Daniel Karavolos}, \bibinfo{person}{Anders Bouwer}, {and} \bibinfo{person}{Rafael Bidarra}.} \bibinfo{year}{2015}\natexlab{}.
\newblock \showarticletitle{{Mixed-initiative design of game levels: integrating mission and space into level generation}}. In \bibinfo{booktitle}{\emph{10th International Conference on the Foundations of Digital Games}}. \bibinfo{publisher}{Foundations of Digital Games}.
\newblock
\urldef\tempurl%
\url{http://graphics.tudelft.nl/Publications-new/2015/KBB15}
\showURL{%
\tempurl}


\bibitem[Karimi et~al\mbox{.}(2020)]%
        {Karimi2020}
\bibfield{author}{\bibinfo{person}{Pegah Karimi}, \bibinfo{person}{Jeba Rezwana}, \bibinfo{person}{Safat Siddiqui}, \bibinfo{person}{Mary~Lou Maher}, {and} \bibinfo{person}{Nasrin Dehbozorgi}.} \bibinfo{year}{2020}\natexlab{}.
\newblock \showarticletitle{{Creative sketching partner}}. In \bibinfo{booktitle}{\emph{Proceedings of the 25th International Conference on Intelligent User Interfaces}}. \bibinfo{publisher}{ACM}, \bibinfo{address}{New York, NY, USA}, \bibinfo{pages}{221--230}.
\newblock
\showISBNx{9781450371186}
\urldef\tempurl%
\url{https://doi.org/10.1145/3377325.3377522}
\showDOI{\tempurl}


\bibitem[Kreminski et~al\mbox{.}(2020)]%
        {Kreminski2020a}
\bibfield{author}{\bibinfo{person}{Max Kreminski}, \bibinfo{person}{Melanie Dickinson}, \bibinfo{person}{Joseph~C. Osborn}, \bibinfo{person}{Adam Summerville}, \bibinfo{person}{Michael Mateas}, {and} \bibinfo{person}{Noah Wardrip-Fruin}.} \bibinfo{year}{2020}\natexlab{}.
\newblock \showarticletitle{{Germinate: A Mixed-Initiative Casual Creator for Rhetorical Games}}. In \bibinfo{booktitle}{\emph{Proceedings of the Sixteenth Artificial Intelligence and Interactive Digital Entertainment Conference}}.
\newblock
\urldef\tempurl%
\url{https://ojs.aaai.org/index.php/AIIDE/article/view/7417}
\showURL{%
\tempurl}


\bibitem[Kreminski et~al\mbox{.}(2022)]%
        {Kreminski2022}
\bibfield{author}{\bibinfo{person}{Max Kreminski}, \bibinfo{person}{Melanie Dickinson}, \bibinfo{person}{Noah Wardrip-Fruin}, {and} \bibinfo{person}{Michael Mateas}.} \bibinfo{year}{2022}\natexlab{}.
\newblock \showarticletitle{{Loose Ends: A Mixed-Initiative Creative Interface for Playful Storytelling}}.
\newblock \bibinfo{journal}{\emph{Proceedings of the AAAI Conference on Artificial Intelligence and Interactive Digital Entertainment}} \bibinfo{volume}{18}, \bibinfo{number}{1} (\bibinfo{date}{oct} \bibinfo{year}{2022}), \bibinfo{pages}{120--128}.
\newblock
\showISSN{2334-0924}
\urldef\tempurl%
\url{https://doi.org/10.1609/aiide.v18i1.21955}
\showDOI{\tempurl}


\bibitem[Lai et~al\mbox{.}(2022)]%
        {Lai2022}
\bibfield{author}{\bibinfo{person}{Gorm Lai}, \bibinfo{person}{Frederic~Fol Leymarie}, {and} \bibinfo{person}{William Latham}.} \bibinfo{year}{2022}\natexlab{}.
\newblock \showarticletitle{{On Mixed-Initiative Content Creation for Video Games}}.
\newblock \bibinfo{journal}{\emph{IEEE Transactions on Games}} \bibinfo{volume}{14}, \bibinfo{number}{4} (\bibinfo{date}{dec} \bibinfo{year}{2022}), \bibinfo{pages}{543--557}.
\newblock
\showISSN{2475-1502}
\urldef\tempurl%
\url{https://doi.org/10.1109/TG.2022.3176215}
\showDOI{\tempurl}


\bibitem[Langendam and Bidarra(2022)]%
        {Langendam2022}
\bibfield{author}{\bibinfo{person}{Thijmen Stefanus~Leendert Langendam} {and} \bibinfo{person}{Rafael Bidarra}.} \bibinfo{year}{2022}\natexlab{}.
\newblock \showarticletitle{{miWFC - Designer Empowerment through mixed-initiative Wave Function Collapse}}. In \bibinfo{booktitle}{\emph{Proceedings of the 17th International Conference on the Foundations of Digital Games}}. \bibinfo{publisher}{ACM}, \bibinfo{address}{New York, NY, USA}, \bibinfo{pages}{1--8}.
\newblock
\showISBNx{9781450397957}
\urldef\tempurl%
\url{https://doi.org/10.1145/3555858.3563266}
\showDOI{\tempurl}


\bibitem[Lee et~al\mbox{.}(2022)]%
        {Lee2022}
\bibfield{author}{\bibinfo{person}{Mina Lee}, \bibinfo{person}{Percy Liang}, {and} \bibinfo{person}{Qian Yang}.} \bibinfo{year}{2022}\natexlab{}.
\newblock \showarticletitle{{CoAuthor: Designing a Human-AI Collaborative Writing Dataset for Exploring Language Model Capabilities}}. In \bibinfo{booktitle}{\emph{CHI Conference on Human Factors in Computing Systems}}. \bibinfo{publisher}{ACM}, \bibinfo{address}{New York, NY, USA}, \bibinfo{pages}{1--19}.
\newblock
\showISBNx{9781450391573}
\urldef\tempurl%
\url{https://doi.org/10.1145/3491102.3502030}
\showDOI{\tempurl}


\bibitem[Liapis et~al\mbox{.}(2016)]%
        {Liapis2016a}
\bibfield{author}{\bibinfo{person}{Antonios Liapis}, \bibinfo{person}{Gillian Smith}, {and} \bibinfo{person}{Noor Shaker}.} \bibinfo{year}{2016}\natexlab{}.
\newblock \showarticletitle{{Mixed-initiative content creation}}.
\newblock \bibinfo{journal}{\emph{Procedural content generation in games}} (\bibinfo{year}{2016}), \bibinfo{pages}{195--214}.
\newblock
\urldef\tempurl%
\url{https://doi.org/10.1007/978-3-319-42716-4_11}
\showDOI{\tempurl}


\bibitem[Liapis et~al\mbox{.}(2012a)]%
        {Liapis2012}
\bibfield{author}{\bibinfo{person}{Antonios Liapis}, \bibinfo{person}{Georgios~N. Yannakakis}, {and} \bibinfo{person}{Julian Togelius}.} \bibinfo{year}{2012}\natexlab{a}.
\newblock \showarticletitle{{Adapting Models of Visual Aesthetics for Personalized Content Creation}}.
\newblock \bibinfo{journal}{\emph{IEEE Transactions on Computational Intelligence and AI in Games}} \bibinfo{volume}{4}, \bibinfo{number}{3} (\bibinfo{date}{sep} \bibinfo{year}{2012}), \bibinfo{pages}{213--228}.
\newblock
\showISSN{1943-068X}
\urldef\tempurl%
\url{https://doi.org/10.1109/TCIAIG.2012.2192438}
\showDOI{\tempurl}


\bibitem[Liapis et~al\mbox{.}(2012b)]%
        {Liapis2012a}
\bibfield{author}{\bibinfo{person}{Antonios Liapis}, \bibinfo{person}{Georgios~N. Yannakakis}, {and} \bibinfo{person}{Julian Togelius}.} \bibinfo{year}{2012}\natexlab{b}.
\newblock \showarticletitle{{Co-creating game content using an adaptive model of user taste}}. In \bibinfo{booktitle}{\emph{Proceedings of the 3rd International Conference on Computational Creativity}}.
\newblock


\bibitem[Liapis et~al\mbox{.}(2013)]%
        {Liapis2013}
\bibfield{author}{\bibinfo{person}{Antonios Liapis}, \bibinfo{person}{Georgios~N. Yannakakis}, {and} \bibinfo{person}{Julian Togelius}.} \bibinfo{year}{2013}\natexlab{}.
\newblock \showarticletitle{{Sentient sketchbook : computer-assisted game level authoring}}. In \bibinfo{booktitle}{\emph{8th International Conference on the Foundations of Digital Games}}. \bibinfo{publisher}{Foundations of Digital Games}, \bibinfo{address}{Chania}.
\newblock


\bibitem[Lin et~al\mbox{.}(2022)]%
        {Lin2022}
\bibfield{author}{\bibinfo{person}{Zhiyu Lin}, \bibinfo{person}{Rohan Agarwal}, {and} \bibinfo{person}{Mark Riedl}.} \bibinfo{year}{2022}\natexlab{}.
\newblock \showarticletitle{{Creative Wand: A System to Study Effects of Communications in Co-creative Settings}}.
\newblock \bibinfo{journal}{\emph{Proceedings of the AAAI Conference on Artificial Intelligence and Interactive Digital Entertainment}} \bibinfo{volume}{18}, \bibinfo{number}{1} (\bibinfo{date}{oct} \bibinfo{year}{2022}), \bibinfo{pages}{45--52}.
\newblock
\showISSN{2334-0924}
\urldef\tempurl%
\url{https://doi.org/10.1609/aiide.v18i1.21946}
\showDOI{\tempurl}


\bibitem[Llano et~al\mbox{.}(2020)]%
        {Llano2020}
\bibfield{author}{\bibinfo{person}{Maria~Teresa Llano}, \bibinfo{person}{Mark D'Inverno}, \bibinfo{person}{Matthew Yee-King}, \bibinfo{person}{Jon McCormack}, \bibinfo{person}{Alon Ilsar}, \bibinfo{person}{Alison Pease}, {and} \bibinfo{person}{Simon Colton}.} \bibinfo{year}{2020}\natexlab{}.
\newblock \showarticletitle{{Explainable Computational Creativity}}. In \bibinfo{booktitle}{\emph{11th International Conference on Computational Creativity}}. \bibinfo{pages}{334--341}.
\newblock


\bibitem[Machado et~al\mbox{.}(2019a)]%
        {Machado2019}
\bibfield{author}{\bibinfo{person}{Tiago Machado}, \bibinfo{person}{Dan Gopstein}, \bibinfo{person}{Andy Nealen}, {and} \bibinfo{person}{Julian Togelius}.} \bibinfo{year}{2019}\natexlab{a}.
\newblock \showarticletitle{{Pitako - Recommending Game Design Elements in Cicero}}. In \bibinfo{booktitle}{\emph{2019 IEEE Conference on Games (CoG)}}. \bibinfo{publisher}{IEEE}, \bibinfo{pages}{1--8}.
\newblock
\showISBNx{978-1-7281-1884-0}
\urldef\tempurl%
\url{https://doi.org/10.1109/CIG.2019.8848081}
\showDOI{\tempurl}


\bibitem[Machado et~al\mbox{.}(2019b)]%
        {Machado2019a}
\bibfield{author}{\bibinfo{person}{Tiago Machado}, \bibinfo{person}{Daniel Gopstein}, \bibinfo{person}{Angela Wang}, \bibinfo{person}{Oded Nov}, \bibinfo{person}{Andrew Nealen}, {and} \bibinfo{person}{Julian Togelius}.} \bibinfo{year}{2019}\natexlab{b}.
\newblock \showarticletitle{{Evaluation of a Recommender System for Assisting Novice Game Designers}}.
\newblock \bibinfo{journal}{\emph{Proceedings of the AAAI Conference on Artificial Intelligence and Interactive Digital Entertainment}} \bibinfo{volume}{15}, \bibinfo{number}{1} (\bibinfo{date}{oct} \bibinfo{year}{2019}), \bibinfo{pages}{167--173}.
\newblock
\showISSN{2334-0924}
\urldef\tempurl%
\url{https://doi.org/10.1609/aiide.v15i1.5240}
\showDOI{\tempurl}


\bibitem[Margarido et~al\mbox{.}(2022)]%
        {Margarido2022}
\bibfield{author}{\bibinfo{person}{Solange Margarido}, \bibinfo{person}{Penousal Machado}, \bibinfo{person}{Licinio Roque}, {and} \bibinfo{person}{Pedro Martins}.} \bibinfo{year}{2022}\natexlab{}.
\newblock \showarticletitle{{Let's Make Games Together: Explainability in Mixed-initiative Co-creative Game Design}}. In \bibinfo{booktitle}{\emph{2022 IEEE Conference on Games (CoG)}}. \bibinfo{publisher}{IEEE}, \bibinfo{pages}{638--645}.
\newblock
\showISBNx{978-1-6654-5989-1}
\urldef\tempurl%
\url{https://doi.org/10.1109/CoG51982.2022.9893636}
\showDOI{\tempurl}


\bibitem[Mobramaein et~al\mbox{.}(2018)]%
        {Mobramaein2018}
\bibfield{author}{\bibinfo{person}{Afshin Mobramaein}, \bibinfo{person}{Morteza Behrooz}, {and} \bibinfo{person}{Jim Whitehead}.} \bibinfo{year}{2018}\natexlab{}.
\newblock \showarticletitle{{CADI — A Conversational Assistive Design Interface for Discovering Pong Variants}}.
\newblock \bibinfo{journal}{\emph{Proceedings of the AAAI Conference on Artificial Intelligence and Interactive Digital Entertainment}} \bibinfo{volume}{14}, \bibinfo{number}{1} (\bibinfo{date}{sep} \bibinfo{year}{2018}), \bibinfo{pages}{194--200}.
\newblock
\showISSN{2334-0924}
\urldef\tempurl%
\url{https://doi.org/10.1609/aiide.v14i1.13042}
\showDOI{\tempurl}


\bibitem[Muller et~al\mbox{.}(2020)]%
        {Muller2020}
\bibfield{author}{\bibinfo{person}{Michael Muller}, \bibinfo{person}{Justin~D Weisz}, {and} \bibinfo{person}{Werner Geyer}.} \bibinfo{year}{2020}\natexlab{}.
\newblock \showarticletitle{{Mixed initiative generative AI interfaces: An analytic framework for generative AI applications}}. In \bibinfo{booktitle}{\emph{Proceedings of the Workshop The Future of Co-Creative Systems - A Workshop on Human-Computer Co-Creativity of the 11th International Conference on Computational Creativity}}.
\newblock


\bibitem[Nelson et~al\mbox{.}(2017)]%
        {Nelson2017}
\bibfield{author}{\bibinfo{person}{Mark~J. Nelson}, \bibinfo{person}{Simon Colton}, \bibinfo{person}{Edward~J. Powley}, \bibinfo{person}{Swen~E. Gaudl}, \bibinfo{person}{Peter Ivey}, \bibinfo{person}{Rob Saunders}, \bibinfo{person}{Blanca~P{\'{e}}rez Ferrer}, {and} \bibinfo{person}{Michael Cook}.} \bibinfo{year}{2017}\natexlab{}.
\newblock \showarticletitle{{Mixed-initiative approaches to on-device mobile game design}}. In \bibinfo{booktitle}{\emph{CHI'17 Workshop on Mixed-Initiative Creative Interfaces}}.
\newblock
\urldef\tempurl%
\url{http://repository.falmouth.ac.uk/id/eprint/2568}
\showURL{%
\tempurl}


\bibitem[Partlan et~al\mbox{.}(2022)]%
        {Partlan2022}
\bibfield{author}{\bibinfo{person}{Nathan Partlan}, \bibinfo{person}{Luis Soto}, \bibinfo{person}{Jim Howe}, \bibinfo{person}{Sarthak Shrivastava}, \bibinfo{person}{Magy {Seif El-Nasr}}, {and} \bibinfo{person}{Stacy Marsella}.} \bibinfo{year}{2022}\natexlab{}.
\newblock \showarticletitle{{EvolvingBehavior: Towards Co-Creative Evolution of Behavior Trees for Game NPCs}}. In \bibinfo{booktitle}{\emph{Proceedings of the 17th International Conference on the Foundations of Digital Games}}. \bibinfo{publisher}{ACM}, \bibinfo{address}{New York, NY, USA}, \bibinfo{pages}{1--13}.
\newblock
\showISBNx{9781450397957}
\urldef\tempurl%
\url{https://doi.org/10.1145/3555858.3555896}
\showDOI{\tempurl}


\bibitem[Rezwana and Maher(2021)]%
        {Rezwana2021}
\bibfield{author}{\bibinfo{person}{Jeba Rezwana} {and} \bibinfo{person}{Mary~Lou Maher}.} \bibinfo{year}{2021}\natexlab{}.
\newblock \showarticletitle{{COFI: A Framework for Modeling Interaction in Human-AI Co-Creative Systems}}. In \bibinfo{booktitle}{\emph{12th International Conference on Computational Creativity}}. \bibinfo{pages}{444--448}.
\newblock


\bibitem[Rezwana and Maher(2023)]%
        {Rezwana2022}
\bibfield{author}{\bibinfo{person}{Jeba Rezwana} {and} \bibinfo{person}{Mary~Lou Maher}.} \bibinfo{year}{2023}\natexlab{}.
\newblock \showarticletitle{{Designing Creative AI Partners with COFI: A Framework for Modeling Interaction in Human-AI Co-Creative Systems}}.
\newblock \bibinfo{journal}{\emph{ACM Transactions on Computer-Human Interaction}} \bibinfo{volume}{30}, \bibinfo{number}{5} (\bibinfo{date}{oct} \bibinfo{year}{2023}), \bibinfo{pages}{1--28}.
\newblock
\showISSN{1073-0516}
\urldef\tempurl%
\url{https://doi.org/10.1145/3519026}
\showDOI{\tempurl}


\bibitem[{Rodriguez Prado} et~al\mbox{.}(2024)]%
        {RodriguezPrado2024}
\bibfield{author}{\bibinfo{person}{Diego {Rodriguez Prado}}, \bibinfo{person}{Victor Perez}, \bibinfo{person}{Mihai Petrescu}, \bibinfo{person}{Titus Ebbecke}, \bibinfo{person}{Erwann Millon}, {and} \bibinfo{person}{Tianpei Gu}.} \bibinfo{year}{2024}\natexlab{}.
\newblock \showarticletitle{{Real-time AI and the Future of Creative Tools}}. In \bibinfo{booktitle}{\emph{ACM SIGGRAPH 2024 Real-Time Live!}} \bibinfo{publisher}{ACM}, \bibinfo{address}{New York, NY, USA}, \bibinfo{pages}{1--2}.
\newblock
\showISBNx{9798400705267}
\urldef\tempurl%
\url{https://doi.org/10.1145/3641520.3665303}
\showDOI{\tempurl}


\bibitem[Samuel et~al\mbox{.}(2016)]%
        {Samuel2016}
\bibfield{author}{\bibinfo{person}{Ben Samuel}, \bibinfo{person}{Michael Mateas}, {and} \bibinfo{person}{Noah Wardrip-Fruin}.} \bibinfo{year}{2016}\natexlab{}.
\newblock \showarticletitle{{The Design of Writing Buddy: A Mixed-Initiative Approach Towards Computational Story Collaboration}}. In \bibinfo{booktitle}{\emph{Interactive Storytelling: 9th International Conference on Interactive Digital Storytellingg, ICIDS 2016}}. \bibinfo{publisher}{Springer International Publishing}, \bibinfo{pages}{388--396}.
\newblock
\urldef\tempurl%
\url{https://doi.org/10.1007/978-3-319-48279-8_34}
\showDOI{\tempurl}


\bibitem[Schell(2008)]%
        {Schell2008}
\bibfield{author}{\bibinfo{person}{Jesse Schell}.} \bibinfo{year}{2008}\natexlab{}.
\newblock \bibinfo{booktitle}{\emph{{The Art of Game Design: A Book of Lenses}}}.
\newblock \bibinfo{publisher}{Morgan Kaufmann Publishers Inc.}
\newblock


\bibitem[Shaker et~al\mbox{.}(2013a)]%
        {Shaker2013a}
\bibfield{author}{\bibinfo{person}{Noor Shaker}, \bibinfo{person}{Mohammad Shaker}, {and} \bibinfo{person}{Julian Togelius}.} \bibinfo{year}{2013}\natexlab{a}.
\newblock \showarticletitle{{Evolving Playable Content for Cut the Rope through a Simulation-Based Approach}}.
\newblock \bibinfo{journal}{\emph{Proceedings of the AAAI Conference on Artificial Intelligence and Interactive Digital Entertainment}} \bibinfo{volume}{9}, \bibinfo{number}{1} (\bibinfo{date}{jun} \bibinfo{year}{2013}), \bibinfo{pages}{72--78}.
\newblock
\showISSN{2334-0924}
\urldef\tempurl%
\url{https://doi.org/10.1609/aiide.v9i1.12690}
\showDOI{\tempurl}


\bibitem[Shaker et~al\mbox{.}(2013b)]%
        {Shaker2013}
\bibfield{author}{\bibinfo{person}{Noor Shaker}, \bibinfo{person}{Mohammad Shaker}, {and} \bibinfo{person}{Julian Togelius}.} \bibinfo{year}{2013}\natexlab{b}.
\newblock \showarticletitle{{Ropossum: An Authoring Tool for Designing, Optimizing and Solving Cut the Rope Levels}}.
\newblock \bibinfo{journal}{\emph{Proceedings of the AAAI Conference on Artificial Intelligence and Interactive Digital Entertainment}} \bibinfo{volume}{9}, \bibinfo{number}{1} (\bibinfo{date}{jun} \bibinfo{year}{2013}), \bibinfo{pages}{215--216}.
\newblock
\showISSN{2334-0924}
\urldef\tempurl%
\url{https://doi.org/10.1609/aiide.v9i1.12611}
\showDOI{\tempurl}


\bibitem[Shaker et~al\mbox{.}(2016)]%
        {Shaker2016}
\bibfield{author}{\bibinfo{person}{Noor Shaker}, \bibinfo{person}{Julian Togelius}, {and} \bibinfo{person}{Mark~J Nelson}.} \bibinfo{year}{2016}\natexlab{}.
\newblock \bibinfo{booktitle}{\emph{{Procedural Content Generation in Games}}}.
\newblock \bibinfo{publisher}{Springer}.
\newblock


\bibitem[Smith et~al\mbox{.}(2010)]%
        {Smith2010}
\bibfield{author}{\bibinfo{person}{Gillian Smith}, \bibinfo{person}{Jim Whitehead}, {and} \bibinfo{person}{Michael Mateas}.} \bibinfo{year}{2010}\natexlab{}.
\newblock \showarticletitle{{Tanagra}}. In \bibinfo{booktitle}{\emph{Proceedings of the Fifth International Conference on the Foundations of Digital Games}}. \bibinfo{publisher}{ACM}, \bibinfo{address}{New York, NY, USA}, \bibinfo{pages}{209--216}.
\newblock
\showISBNx{9781605589374}
\urldef\tempurl%
\url{https://doi.org/10.1145/1822348.1822376}
\showDOI{\tempurl}


\bibitem[Smith et~al\mbox{.}(2011)]%
        {Smith2011}
\bibfield{author}{\bibinfo{person}{Gillian Smith}, \bibinfo{person}{Jim Whitehead}, {and} \bibinfo{person}{Michael Mateas}.} \bibinfo{year}{2011}\natexlab{}.
\newblock \showarticletitle{{Tanagra: Reactive Planning and Constraint Solving for Mixed-Initiative Level Design}}.
\newblock \bibinfo{journal}{\emph{IEEE Transactions on Computational Intelligence and AI in Games}} \bibinfo{volume}{3}, \bibinfo{number}{3} (\bibinfo{date}{sep} \bibinfo{year}{2011}), \bibinfo{pages}{201--215}.
\newblock
\showISSN{1943-068X}
\urldef\tempurl%
\url{https://doi.org/10.1109/TCIAIG.2011.2159716}
\showDOI{\tempurl}


\bibitem[Spoto and Oleynik(2017)]%
        {Spoto2017}
\bibfield{author}{\bibinfo{person}{Angie Spoto} {and} \bibinfo{person}{Natalia Oleynik}.} \bibinfo{year}{2017}\natexlab{}.
\newblock \bibinfo{title}{{Library of Mixed-Initiative Creative Interfaces}}.
\newblock
\newblock
\urldef\tempurl%
\url{http://mici.codingconduct.cc/}
\showURL{%
\tempurl}


\bibitem[Stefnisson and Thue(2018)]%
        {Stefnisson2018}
\bibfield{author}{\bibinfo{person}{Ingibergur Stefnisson} {and} \bibinfo{person}{David Thue}.} \bibinfo{year}{2018}\natexlab{}.
\newblock \showarticletitle{{Mimisbrunnur: AI-Assisted Authoring for Interactive Storytelling}}.
\newblock \bibinfo{journal}{\emph{Proceedings of the AAAI Conference on Artificial Intelligence and Interactive Digital Entertainment}} \bibinfo{volume}{14}, \bibinfo{number}{1} (\bibinfo{date}{sep} \bibinfo{year}{2018}), \bibinfo{pages}{236--242}.
\newblock
\showISSN{2334-0924}
\urldef\tempurl%
\url{https://doi.org/10.1609/aiide.v14i1.13046}
\showDOI{\tempurl}


\bibitem[Sun et~al\mbox{.}(2022)]%
        {Sun2022}
\bibfield{author}{\bibinfo{person}{Yuqian Sun}, \bibinfo{person}{Xuran Ni}, \bibinfo{person}{Haozhen Feng}, \bibinfo{person}{Ray LC}, \bibinfo{person}{Chang~Hee Lee}, {and} \bibinfo{person}{Ali Asadipour}.} \bibinfo{year}{2022}\natexlab{}.
\newblock \showarticletitle{{Bringing stories to life in 1001 nights: A co- creative text adventure game using a story generation model}}. In \bibinfo{booktitle}{\emph{International Conference on Interactive Digital Storytelling}}. \bibinfo{publisher}{Springer}, \bibinfo{pages}{651----672}.
\newblock


\bibitem[Triyason(2023)]%
        {Triyason2023}
\bibfield{author}{\bibinfo{person}{Tuul Triyason}.} \bibinfo{year}{2023}\natexlab{}.
\newblock \showarticletitle{{Exploring the Potential of ChatGPT as a Dungeon Master in Dungeons \& Dragons tabletop game}}. In \bibinfo{booktitle}{\emph{Proceedings of the 13th International Conference on Advances in Information Technology}}. \bibinfo{publisher}{ACM}, \bibinfo{address}{New York, NY, USA}, \bibinfo{pages}{1--6}.
\newblock
\showISBNx{9798400708497}
\urldef\tempurl%
\url{https://doi.org/10.1145/3628454.3628457}
\showDOI{\tempurl}


\bibitem[Walton et~al\mbox{.}(2022)]%
        {Walton2020}
\bibfield{author}{\bibinfo{person}{Sean~P. Walton}, \bibinfo{person}{Alma A.~M. Rahat}, {and} \bibinfo{person}{James Stovold}.} \bibinfo{year}{2022}\natexlab{}.
\newblock \showarticletitle{{Evaluating Mixed-Initiative Procedural Level Design Tools Using a Triple-Blind Mixed-Method User Study}}.
\newblock \bibinfo{journal}{\emph{IEEE Transactions on Games}} \bibinfo{volume}{14}, \bibinfo{number}{3} (\bibinfo{date}{sep} \bibinfo{year}{2022}), \bibinfo{pages}{413--422}.
\newblock
\showISSN{2475-1502}
\urldef\tempurl%
\url{https://doi.org/10.1109/TG.2021.3086215}
\showDOI{\tempurl}


\bibitem[Yannakakis et~al\mbox{.}(2014)]%
        {Yannakakis2014}
\bibfield{author}{\bibinfo{person}{Georgios~N. Yannakakis}, \bibinfo{person}{Antonios Liapis}, {and} \bibinfo{person}{Constantine Alexopoulos}.} \bibinfo{year}{2014}\natexlab{}.
\newblock \showarticletitle{{Mixed-initiative co-creativity}}. In \bibinfo{booktitle}{\emph{9th International Conference on the Foundations of Digital Games}}. \bibinfo{publisher}{Foundations of Digital Games}.
\newblock
\urldef\tempurl%
\url{https://www.um.edu.mt/library/oar/handle/123456789/29459}
\showURL{%
\tempurl}


\bibitem[Yuan et~al\mbox{.}(2022)]%
        {Yuan2022}
\bibfield{author}{\bibinfo{person}{Ann Yuan}, \bibinfo{person}{Andy Coenen}, \bibinfo{person}{Emily Reif}, {and} \bibinfo{person}{Daphne Ippolito}.} \bibinfo{year}{2022}\natexlab{}.
\newblock \showarticletitle{{Wordcraft: Story Writing With Large Language Models}}. In \bibinfo{booktitle}{\emph{27th International Conference on Intelligent User Interfaces}}. \bibinfo{publisher}{ACM}, \bibinfo{address}{New York, NY, USA}, \bibinfo{pages}{841--852}.
\newblock
\showISBNx{9781450391443}
\urldef\tempurl%
\url{https://doi.org/10.1145/3490099.3511105}
\showDOI{\tempurl}


\bibitem[Zhu et~al\mbox{.}(2018)]%
        {Zhu2018}
\bibfield{author}{\bibinfo{person}{Jichen Zhu}, \bibinfo{person}{Antonios Liapis}, \bibinfo{person}{Sebastian Risi}, \bibinfo{person}{Rafael Bidarra}, {and} \bibinfo{person}{G.~Michael Youngblood}.} \bibinfo{year}{2018}\natexlab{}.
\newblock \showarticletitle{{Explainable AI for Designers: A Human-Centered Perspective on Mixed-Initiative Co-Creation}}. In \bibinfo{booktitle}{\emph{2018 IEEE Conference on Computational Intelligence and Games (CIG)}}. \bibinfo{publisher}{IEEE}, \bibinfo{pages}{1--8}.
\newblock
\showISBNx{978-1-5386-4359-4}
\urldef\tempurl%
\url{https://doi.org/10.1109/CIG.2018.8490433}
\showDOI{\tempurl}


\end{thebibliography}


\end{document}